\documentclass[final,5p,times,twocolumn]{elsarticle}

\journal{Automatica}

\makeatletter
\def\ps@pprintTitle{%
	\let\@oddhead\@empty
	\let\@evenhead\@empty
	\let\@oddfoot\@empty
	\let\@evenfoot\@oddfoot
}
\makeatother

\usepackage{wrapfig}
\usepackage{graphicx}
\usepackage{graphics} 
\usepackage{epsfig} 
\usepackage{mathptmx} 
\usepackage{times} 
\usepackage{amsmath} 
\usepackage{amssymb}  
\usepackage{caption}
\usepackage{subcaption}

\usepackage{enumitem}

\usepackage[dvipsnames]{color}
\usepackage{tikz}

\usetikzlibrary{shapes,arrows}
\usetikzlibrary{positioning}
\usetikzlibrary{arrows.meta}
\usepackage{pgfplots}
\pgfplotsset{compat=newest} 
\usepackage{pgfplotstable}
\usepackage{lipsum}  

\RequirePackage{luatex85}

\UseRawInputEncoding

\makeatletter

\def\endthebibliography{%
	\def\@noitemerr{\@latex@warning{Empty `thebibliography' environment}}%
	\endlist
}
\makeatother

\newcommand{\R}{\mathbb{R}}

\newcommand{\C}{\mathcal{C}}

\newtheorem{theorem}{Theorem}

\newtheorem{assumption}{Assumption}

\newtheorem{proposition}{Proposition}

\newtheorem{definition}{Definition}

\newcommand{\norm}[1]{\left\lVert#1\right\rVert}
\newcommand{\abs}[1]{\left\lvert#1\right\rvert}

\definecolor{mypink}{cmyk}{0, 0.7808, 0.4429, 0.1412}

\begin{document}
	
	\tikzstyle{block} = [draw, rectangle, 
	minimum height=3em]
	\tikzstyle{sum} = [draw, circle, scale = 0.5,node distance = 1cm]
	\tikzstyle{product} = [draw, circle, cross, minimum width=1 cm, scale = 0.5]
	\tikzstyle{input} = [coordinate]
	\tikzstyle{output} = [coordinate]
	\tikzstyle{pinstyle} = [pin edge={to-,thin,black}]
	\tikzstyle{gain} gain  = [draw, thick, isosceles triangle, minimum height = 2em, isosceles triangle apex angle=60, scale = 0.8]
	
	\setlength{\paperheight}{297mm}
	\setlength{\paperwidth}{210mm}
	
	\thispagestyle{empty}
	\bibliographystyle{IEEEtranS}
	
	\title{\LARGE{\bf
			Practical Prescribed-Time Seeking of a Repulsive Source \\ 
			by Unicycle Angular Velocity Tuning}
	}
	
	\author[author1]{Velimir Todorovski}\ead{velimir.todorovski@tum.de} 
	\author[author2]{Miroslav Krstic}\ead{krstic@ucsd.edu}
	
	\address[author1]{
		Technical University of Munich, Munich, Germany}
	\address[author2]{Department of Mechanical and Aerospace Engineering, University of California, San Diego, USA}


	\begin{abstract}

		All the existing source seeking algorithms for unicycle models in GPS-denied settings guarantee at best an exponential rate of convergence over an infinite interval. Using the recently introduced time-varying feedback tools for prescribed-time stabilization, we achieve source seeking in prescribed time, i.e., the convergence to a small but bounded neighborhood of the source, without the measurements of the position and velocity of the unicycle, in as short a time as the user desires, starting from an arbitrary distance from the source. The practical convergence is established using a singularly perturbed version of the Lie bracket averaging method, combined with time dilation and time contraction operations. The algorithm is robust, provably, even to bounded nonvanishing perturbations and an arbitrarily strong gradient-dependent repulsive velocity drift emanating from the source.
	\end{abstract}
	
	\date{}
	\maketitle
	

	
	\section{Introduction}
	
	Source seeking has received  considerable  attention in  recent years due to its application in navigation of autonomous agents in GPS-denied environments. The goal of the source seeking problem is to guide a vehicle to the
	location of a source that emits some type of a measurable signal. As there is no
	GPS data available, this should be achieved without the knowledge of the position
	of the vehicle. Furthermore, the signal radiated from the source forms an unknown
	scalar value field in space. This could be a concentration of a chemical or biological
	agent, but also an electromagnetic, acoustic, thermal or radar signal \cite{vapor_emit}. It
	is assumed, that the vehicle is equipped with a sensor that measures the value of the signal field at its location. The strength of the signal has its maximum value at the source and decays away from it through diffusion or other physical processes. 
	As a consequence of this, source seeking can be thought of as an optimization problem of finding the maximum value of a scalar field. In addition, the agent must be steered to the source without using its position information since it is assumed that it operates in an environment where no GPS data available. The simultaneous solution of the model-free steering and optimization tasks is possible with the extremum seeking (ES) method \cite{krstic2000stability}.
	In addition to guidance of autonomous vehicles, the topic of source seeking has a variety of other promising applications, including wireless communication, search and rescue operations, medical science and security engineering  \cite{application}.

	This paper addresses the problem of seeking a {\em potentially repulsive} source for a velocity actuated unicycle vehicle model. Additionally, the source is sought in finite user-prescribed time (PT). This is achieved by employing an  existing ES-based source seeking scheme and enhancing
	it with tools recently developed for prescribed-time stabilization. These tools build
	on classical proportional navigation laws for tactical missile guidance \cite{missile}, an
	application in which the control objective---if achieved---is  achieved in finite time that is independent of the initial conditions (of the angle relative to the light of sight, and the associated angular velocity). 

	\textit{1.1. Motivation:} Scheinker and Krsti\'{c} examine the question of model-free stabilization by ES in \cite{scheinker2017model}. Herein, ES control laws are designed for a class of nonlinear control affine systems with possibly unstable drift terms.
		In addition to stabilizing an unstable plant, the developed ES controllers can be used for position deprived source seeking.
		However, one disadvantage of the proposed controllers is that
		their stabilization abilities are dependent on the initial conditions and the choice of appropriate gains. In the context of source seeking this implies that the source is located only if these criteria are satisfied.
		In this manuscript an effort to overcome this limitation is provided for two classes of second-order unstable drift terms. The repulsive source is reached regardless of initial conditions and the choice of gains. Additionally, this is achieved in predefined finite time. This feature is of practical importance and it is suitable for applications where time is
		of the essence. For instance, consider the scenario described in \cite{SandR} where a mountain skier is buried in snow by an avalanche. It is assumed, that the skier is equipped with a device that emits an electromagnetic field in space. Reaching the source of the signal in finite time would dramatically increase the chances of survival of the person trapped under snow. 
	
\textit{1.2. Literature review:}	The related works to this contribution stem from two fields. Namely, source seeking via ES and PT stabilization of systems. 
	The first ES based controller for source seeking was developed in~\cite{first_ss} for a velocity actuated point mass. The source seeking problem was solved by exploiting the integrator dynamics of the vehicle model combined with an
	ES scheme. Additionally, Zhang et al. \cite{zhang2007source} consider a more realistic model of the vehicle, namely, a unicyle. In here, the forward velocity of the unicycle was tuned with
	an ES feedback law which moves the unicycle towards a bounded region around the
	source. 
	The ES based control law
	from \cite{zhang2007source}, however, can produce trajectories that have a triangular or diamond
	shaped form that are unfeasible for some vehicles, such as a fixed-wing aircraft
	that can not move backwards. To address this, Cochran and Krsti\'{c} \cite{cochran2009nonholonomic} proposed
	a source seeking scheme that tunes the angular velocity of the unicycle by explicitly
	injecting sinusoidal perturbations. The resulting trajectories are more realistic and
	applicable to a broader range of vehicles.  Contrary to this approach, in \cite{ES_with_bounded_update_rates} the
		tuning of the angular velocity was achieved by exploiting the special structure of the
		unicycle dynamics. The advantage of this method is that the convergence rate to
		the source and the control effort for the unicycle inputs remain bounded, which
		is important for actual hardware implementation. One drawback of the
		method presented in \cite{ES_with_bounded_update_rates} is that it requires direct actuation of the heading angle
		of the unicycle which in most cases is not possible. An extension of this method was
		presented in \cite{ES_using_LBAandSP}, where the sensor measurement is filtered with an approximative differentiator. This improves the transient behavior of the unicycle to the
		source and more importantly allows for the source seeking scheme from \cite{ES_with_bounded_update_rates} to
		be realized by actuating the angular velocity. This result is the foundation on which our main design from Section \ref{sec:design} is developed. For
	several applications, controlling the angular acceleration is more realistic than a
	direct control of the angular velocity. In this direction, Suttner proposed an ES
	based source seeking scheme with acceleration actuated unicycle in \cite{suttner2019extremum, suttner2020acceleration} based on symmetric product approximations. These control laws have the advantage that
	the angular velocity remains bounded at high frequencies. 
	It is important to note that all ES based source seeking schemes exhibit practical asymptotic stability properties, i.e., the vehicles converge to a small unknown but bounded neighborhood around the source as time goes to infinity. 	
On a different note, the problem of PT stabilization requires that the terminal time $T$, in which the system is stabilized, is a parameter that can be prescribed by the control designer and that it is independent of initial conditions. The first effort to address this, was by Song et al.~\cite{song2017time} where PT regulation of a nonlinear system in normal form was achieved by scaling the states with time-varying gains that grow
	unbounded as the terminal time is reached. The controller is designed for the scaled
	state system which results in PT regulation of the original states. The proposed PT
	controllers were proven to be robust against matched uncertainties which highlighted the potential
	advantage of using PT control laws. 
	An extension of the state
	scaling approach was proposed by Holloway et al.~\cite{holloway2019PTobservers,holloway2019PTobservers_2}
	for PT observers and PT output feedback controllers for linear systems  that converge in finite time. 
	This method allowed for both easy prescription of the convergence times,
	and minimal tuning of the observer and controller parameters. 
	More generally, the PT-convergence and-stabilization of uncertain nonlinear systems was investigated in \cite{krishna_3,krishna_2} by introducing the temporal transformations approach. Contrary to the state scaling approach from above,
	this approach has the advantage of using existing stability analysis techniques to
	examine the convergence of the designed PT control systems. The developed framework for
	design and analysis of PT control laws based on the temporal transformations is
	adopted in this work and it is revised in the preliminary Section
	\ref{ssec:temporal_transformations}.
	
\textit{1.3. Contribution:} We present a novel source seeking algorithm named PT Seeker which tunes the angular velocity to find the source of an unknown scalar signal in a user-prescribed finite time. This represents a major advancement relative to the previous asymptotic/infinite-time seeking algorithms. In addition to tuning the angular velocity, both the angular and  forward velocities are scaled by a time-varying gain which grows unbounded as the terminal time is reached. The use of time-varying PT feedback endows the PT Seeker with robustness to bounded external disturbances and a potentially destabilizing velocity drift term emanated from the source---a first such result in source seeking. We prove convergence to the source in prescribed time regardless of initial conditions and the choice of gains. This is achieved by employing the temporal transformations. In particular, the time dilation which allows us to use an already existing technique for analyzing ES systems, i.e., the singularly perturbed Lie bracket analysis and the time contraction with which our PT Seeker achieves the newly introduced stability property FxT-sSPUAS (see Def. 4).
To avoid numerical instability, in practice we bound the time-varying gains to some user defined maximum values. Nevertheless, our simulation results show that the PT convergence speed and the robustness to the drift are \textit{pratically} preserved.
This paper is a journal version of \cite{TodKrs-ACC22} and contains, in addition, a tutorial on the temporal transformations approach, a generalization from quadratic maps $F$ to maps that are bounded by even polynomias, proof of robustness of our algorithm to bounded nonvanishing perturbations and a comparison study on the use of bounded time-varying gains versus constant high gains.   

	\textit{1.4.~Organization and notation:} In Section \ref{sec:preliminaries}, we review the singularly perturbed Lie bracket approximations method and introduce an approach for design and analysis of PT controllers via temporal transformations. In Section \ref{sec:problem_statemenet} we formulate the problem of PT seeking of a repulsive source  and in Section \ref{sec:design} we systematically derive the PT Seeker. 
	The analysis of the convergence of the PT Seeker to the source is done in Section \ref{sec:convergence_analysis}. Furthermore, we discuss feasibility and advantages of our design in Section \ref{sec:feasibility}. We conclude by presenting our simulation results in Section \ref{sec:simulation}. The Euclidian norm is denoted as $\norm{\cdot}$ and class $\mathcal{K}$, $\mathcal{K}_{\infty}$ and $\mathcal{KL}$ functions are defined as in \cite{khalil2002nonlinear}. The rest of the notation is in accordance with  D\"{u}rr et al. \cite{LBA_first}. 
	

	\section{Preliminaries}
	\label{sec:preliminaries}
	\textit{2.1. Singularly perturbed Lie bracket approximations: \label{ssec:LBA_SP}}
	In the following, we revise the method for stability analysis of extremum seeking systems established in \cite{ Singularly_Pertubed_LBA, ES_using_LBAandSP}. This method is based upon the combination of Singular Pertubations \cite[Chapter 11]{khalil2002nonlinear} and the Lie Bracket Approximations \cite{LBA_first}.
	Consider a system of the form 
	\begin{subequations}
		\label{eq:general_nominal_system}
		\begin{align}
		\frac{\text{d} x }{\text{d} \tau}&= \mu b_0(\bar{\tau}, x, z)  + \mu \sqrt{\omega} \sum\limits_{i=1}^{N}b_i(\bar{\tau}, x, z)u_i(\bar{\tau},\omega \bar{\tau}) \\
		\frac{\text{d} z }{\text{d} \tau} &= g(x,z) 
		\end{align}
	\end{subequations}
	with $\tau := [t_0,\infty)$, the initial conditions $x(t_0) = x_0 \in \mathbb{R}^n$, $z(t_0) = z_0 \in \mathbb{R}^m$, where $n,m \in \mathbb{N}$, the parameters $\mu, \omega \in \left(0, \infty \right)$, $\bar{\tau} = \mu \tau$ and $N \in \mathbb{N}$. Moreover, suppose that the next assumption holds. 
		\begin{assumption}[\cite{Singularly_Pertubed_LBA,ES_using_LBAandSP}]
			\label{ass:singular_ass}
			System \eqref{eq:general_nominal_system} satisfies the following:
			\begin{enumerate}[label=\Alph*.,ref=\ref{ass:singular_ass}\Alph*]
				\item  $b_i \in \C^2 : \mathbb{R} \times \mathbb{R}^n \times \mathbb{R}^m \rightarrow \mathbb{R}^n$ for $i = 0,\ldots,N$.
				\item The vector fields $b_i$, $i = 0,\ldots,N$ are bounded in their first argument up to the second derivative, i.e., for all compact sets $\mathcal{C}_x \subseteq \mathbb{R}^n$ and $ \mathcal{C}_z \subseteq \mathbb{R}^m$ there exist $A_1, \ldots, A_6 \in [0, \infty)$ such that $\norm{b_i(\tau,x,z) }\le A_1$, $\norm{\frac{\partial b_i(\tau,x,z)}{\partial \tau} }\le A_2$, $\norm{\frac{\partial b_i(\tau,x,z)}{\partial x}} + \norm{\frac{\partial b_i(\tau,x,z)}{\partial z}} \le A_3$, $\norm{\frac{\partial^2 b_i(\tau,x,z)}{\partial \tau \partial x}} + \norm{\frac{\partial^2 b_i(\tau,x,z)}{\partial t\partial z}} \le A_4$, $\norm{\frac{\partial [b_j,b_k](\tau,x,z)}{\partial x}} + \norm{\frac{\partial [b_j,b_k](\tau,x,z)}{\partial z}} \le A_5$, $\norm{\frac{\partial [b_j,b_k](\tau,x,z)}{\partial \tau}} \le A_6$ for all $x \in \mathcal{C}_x$, $z \in \mathcal{C}_z$, $\tau \in \mathbb{R}$, $i = 0,\ldots,N$, $j = 1,\ldots, N$, $k = j,\ldots,N$. 
				\item $u_i : \R \times \R \rightarrow \R$, $i = 1, \ldots,N$ are Lebesgue-measurable functions. Moreover, for all $i = 1, \ldots,N$ there exist constants $L_i, M_i \in (0, \infty)$ such that $\norm{u_i(\tau_1,\theta) - u_i(\tau_2,\theta)} \le L_i \abs{\tau_1 - \tau_2}$ for all $\tau_1, \tau_2 \in \R$ and such that $\sup \limits_{\tau, \theta \in \R} \norm{u_i(\tau,\theta)} \le M_i$.
				\item $u_i(\tau,\cdot)$ is $T_f$-periodic, i.e., $u_i(\tau, \theta + T_f) = u_i(\tau, \theta)$ and has zero average, i.e., $\int_{0}^{T_f}u_i(\tau,s)\text{d}s = 0$, with $T_f \in (0, \infty)$ for  all $\tau, \theta \in \R$, $i = 1,\ldots,N$.
				
				\item $g \in \mathcal{C}: \R^n \times \R^m \rightarrow \R^m$ is locally Lipschitz.
				\item  There exists a unique $l \in \C^2: \R^n \rightarrow \R^m$ such that $z = l(x) \Leftrightarrow 0 = g(x,z)$ for all $x \in \R^n$.
			\end{enumerate}
		\end{assumption}
	The system \eqref{eq:general_nominal_system} is in a form where the singular pertubations method can be applied. For this, we calculate the so-called quasi-steady-state $l(x)$, that satisfies $g(x,l(x)) = 0$ and perform the change of variables $\tilde{y} = z-l(x)$. 
	By letting $\mu \rightarrow 0$, we obtain the so-called \textit{boundary layer} model as
	\begin{align}
	\frac{\text{d} \tilde{y} }{\text{d} \tau} = g(x, \tilde{y} + l(x)) \label{eq:boundary_layer_model}
	\end{align} with $\tilde{y}(t_0) = \tilde{y}_0 = z_0 - l(x_0) \in \mathbb{R}^m$.
	Imposing $z = l(x)$, i.e., $ \tilde{y} = 0$ the so-called \textit{reduced} model is obtained as
	\begin{equation}
	\frac{\text{d} \tilde{x} }{\text{d} \tau} = \begin{aligned}[t]
	\mu  b_0(\bar{\tau}, \tilde{x}, l(\tilde{x}))  +  \mu \sqrt{w} \sum\limits_{i=1}^{N}b_i(\bar{\tau}, \tilde{x}, l(\tilde{x} ))u_i(\bar{\tau}, \omega \bar{\tau})
	\end{aligned}  \label{eq:reduced_model}
	\end{equation} with $\tilde{x}(t_0) = x_0 \in \mathbb{R}^n$.
	The special form of \eqref{eq:reduced_model} allows for the use of the Lie bracket approximation method, which leads to the Lie bracket system for $\mu = 1$
	\begin{align}
	\frac{\text{d} \bar{x} }{\text{d} \tau} = b_0(\tau, \bar{x},l(\bar{x})) + \sum\limits_{\substack{i = 1 \\ j= i+1}}^{N} \left[b_i, b_j \right](\tau,\bar{x},l(\bar{x}))v_{ji} \label{eq:Lie_Bracket_System}
	\end{align}
	with $\bar{x}(t_0) = x_0$, $v_{ji} = \frac{1}{T} \int_{0}^{T} u_j(\theta) \int_{0}^{\theta} u_i(s) \text{d} s \text{d} \theta$, where $\left[b_1, b_2 \right]$ represents the Lie bracket of the vector fields $b_1$ and $b_2$.  Next, the relationship between the stability properties of \eqref{eq:general_nominal_system}, \eqref{eq:boundary_layer_model} and \eqref{eq:Lie_Bracket_System} is summarized in the following theorem. 
	
	\begin{theorem}[\cite{Singularly_Pertubed_LBA,ES_using_LBAandSP}]
		\label{thm:SPLBave}
		Consider system \eqref{eq:general_nominal_system} and suppose Assumption \ref{ass:singular_ass} is satisfied. Furthermore, suppose that a compact set $\mathcal{S}$ is globaly uniformly asymptotically stable (GUAS) for \eqref{eq:Lie_Bracket_System} and there exist $\mathcal{K}_{\infty}$-functions $\alpha_1$, $\alpha_2$ and a $\mathcal{K}$-function $\alpha_3$: $[0,\infty) \rightarrow [0,\infty) $  and a function $V \in \mathcal{C}: \mathbb{R}^m \rightarrow \mathbb{R} $ such that for all $[x^{\top}, \tilde{y}^{\top}]^{\top} \in \mathbb{R}^n \times \mathbb{R}^m$ 
		\begin{align}
		\alpha_1(\abs{\tilde{y}}) \le V(\tilde{y}) &\le \alpha_2(\abs{\tilde{y}})  \label{eq:blm_req_1}\\
		\frac{\partial V(\tilde{y} )}{\partial \tilde{y}} g(x, \tilde{y}+l(x)) &\le -\alpha_3(\abs{\tilde{y}}) \label{eq:blm_req_2}
		\end{align} are satisfied. Then, the set $\mathcal{S}$ is sSPUAS for \eqref{eq:general_nominal_system}.
	\end{theorem}
The corresponding sSPUAS stability property is found in Definition \ref{def:FxT-sSPUAS} in the Appendix with $T$ set to $\infty$.

	\textit{2.2 Control laws via temporal transformation: 	\label{ssec:temporal_transformations}} The main idea for the design and analysis of control laws by using temporal scale transformations is established by Krishnamurthy et al.~in  \cite{krishna_3,krishna_2}. 
	The objective of this approach is to develop a PT control algorithm over the PT time interval where $t \in I_t := [t_0, t_0 + T)$, so that  the states converge to their steady-state values as $t \rightarrow t_0 + T$, where $T>0$ is prescribed by the user and $t_0\geq 0$. In order to simplify this task, a time scale transformation $a$ is introduced, that maps the finite time interval $I_t$ to the infinite time interval $I_{\tau} := [t_0, \infty)$ in time $\tau \in I_{\tau}$, i.e., $a: I_t \rightarrow I_{\tau}$. Thus, the control objective in time $\tau$ becomes the convergence of the states to their steady-state values as $\tau \rightarrow \infty$ and the control algorithm is designed on $I_{\tau}$. The desired PT-convergence is then achieved by introducing the inverse transformation $a^{-1}:I_{\tau} \rightarrow I_t$ and converting the control algorithm constructed in time $\tau$ to time $t$.
	To illustrate this, consider a simple scalar example as
	\begin{align}
	\frac{\text{d} x}{\text{d} t} = u(t) \label{eq:nominal_example_t}
	\end{align} 
	where $x(t_0) = x_0 \in \R$. The goal is to choose $u(t) \in \R$, such that, $x$ converges to $0$ as $t \rightarrow t_0 + T$. In order to examine \eqref{eq:nominal_example_t} in time $\tau$, we introduce the transformation $a: I_t \rightarrow I_{\tau} $ as
	\begin{equation}
	\tau = t_0 + \frac{t-t_0}{\nu(t-t_0)} \label{eq:tau_to_t} 
	\end{equation}
	where
	\begin{align}
	\nu(t-t_0)  = 1 - \frac{t - t_0}{T} \label{eq:nu}
	\end{align} 
	is a monotonically decreasing linear function with the properties that $v(0) = 1$ and $v(T) = 0$. The transformation \eqref{eq:tau_to_t} has the effect of \textit{dilation} of the finite time interval $I_t$ to the infinite time interval $I_\tau$, which is easily seen by $\lim\limits_{t \rightarrow t_0 + T} \tau = t_0 + \frac{T}{\nu(T)} = \infty$. Considering \eqref{eq:tau_to_t} along with
	\begin{equation}
	\frac{\text{d} t}{\text{d} \tau} = \frac{1}{(1 + \frac{\tau - t_0}{T})^2}, \label{eq:dt_dtau}
	\end{equation}
	we perform the change of variables $t \rightarrow \tau$ and obtain
	\begin{align}
	\frac{\text{d} x}{\text{d} t} \frac{\text{d} t}{\text{d} \tau} = \frac{\text{d} x}{\text{d} \tau} = \frac{1}{(1 + \frac{\tau - t_0}{T})^2} u(\tau). \label{eq:nominal_example_tau}
	\end{align}
	The problem is now transformed into designing a control law $u(\tau)$, such that $x$ converges to $0$ as $\tau \rightarrow \infty$. This is easily achieved by choosing
	\begin{align}
	u(\tau) = - \left(1 + \frac{\tau - t_0}{T}\right)^2 k x \label{eq:u_tau}
	\end{align} which substituted in \eqref{eq:nominal_example_tau} yields
	\begin{align}
	\frac{\text{d} x}{\text{d} \tau} = -k x. \label{eq:example}
	\end{align}
	The linear time-invariant model \eqref{eq:example} is globally uniformly asymptotically stable (GUAS) for $k > 0$ and  $x(\tau)$ converges exponentially to $0$ as $\tau \rightarrow \infty$, which is  evident from the solution 
	\begin{equation}
	x(\tau) = x_0 {\rm e}^{-k(\tau-t_0)} \label{eq:example_solution}.
	\end{equation}
	Since the original problem is posed on the finite time interval  $I_t$, we convert \eqref{eq:u_tau} and \eqref{eq:example} in time $t$ with the inverse transformation $a^{-1}:I_{\tau} \rightarrow I_t$ defined as
	\begin{equation}
	t = t_0 + \frac{\tau  - t_0}{1+\frac{\tau - t_0}{T}}, \label{eq:t_to_tau}   
	\end{equation}
	which amounts to a \textit{contraction} of the infinite time interval $I_{\tau}$ to the finite time interval $I_t$, also apparent from $\lim\limits_{\tau \rightarrow \infty} t = t_0 + (1 - \frac{t_0}{\tau})/(\frac{1}{T} + \frac{1}{\tau}(1-t_0/T) )= t_0 + T$. In view of
	\begin{equation}
	\frac{\text{d} \tau}{\text{d} t} =  \frac{1 }{\nu^2(t-t_0)}, \label{eq:dtau_dt}
	\end{equation}
	we perform the change of variables $\tau \rightarrow t$ and convert  \eqref{eq:example} to 
	\begin{align}
	\frac{ {\rm d} x }{{\rm d} \tau} \frac{ {\rm d} \tau }{{\rm d} t} = \frac{ {\rm d} x }{{\rm d} t} = - \frac{1}{\nu^2(t-t_0)} k x .\label{eq:pt_example}
	\end{align}
	Thus, we recognize 
	\begin{align}
	u(t) = - \frac{1}{\nu^2(t-t_0)} k x \label{eq:u_t}.
	\end{align}
	The controller \eqref{eq:u_t} drives $x(t)$ to $0$ in prescribed time $T$, which is obvious considering the solution of \eqref{eq:pt_example} given by 
	\begin{equation}
	x(t) = x_0 {\rm e}^{-k\frac{t-t_0}{\nu(t-t_0)}}. \label{eq:pt_example_solution}
	\end{equation}
	and $ \lim\limits_{t \rightarrow t_0 +  T } x(t) = x_0 {\rm e}^{-\infty} = 0$.
	Since  \eqref{eq:example_solution} and \eqref{eq:pt_example_solution} refer to the value of the same signal, only in different time scales $\tau$ and $t$, we  conclude that they have the same convergence properties for $\tau \rightarrow \infty$ and $t \rightarrow t_0 + T$, respectively. Hence, as a consequence of \eqref{eq:example} being GUAS on the infinite time interval $I_{\tau}$, the desired GUAS property of \eqref{eq:pt_example} on the finite time interval  $I_t$ is achieved. This property is also referred to as FxT-GUAS and its definition is found in \cite[Def. 1]{holloway2019PTobservers_2}.

\section{Problem Statement}
	\label{sec:problem_statemenet}
	Consider an autonomous vehicle modeled as a nonholonomic unicycle that is placed within an unknown field $F$. Additionally, the position dynamics are subject to an added potentially destabilizing drift term. The equations of motion of the vehicle in this case are  
	\begin{subequations}
		\label{eq:unicycle_with_drift}
		\begin{align} 
		\dot{x} &=  f(t,x) + u_1\begin{bmatrix}
			\cos(\theta) \\
			\sin(\theta)
			\end{bmatrix} 
		\label{eq:unicycle_with_drift-a}
		\\
		\dot{\theta} &= u_2 \\
		y  &= F(x)
		\end{align}
	\end{subequations}
	where $x = [x_1,x_2]^{\top} \in \R^2$ are the coordinates of the vehicles center with  $x(t_0) = x_0$, $\theta \in \R$ is the orientation of the unicycle with $\theta(t_0) = \theta_0$ and $u_1(t), u_2(t) \in \R$ are the forward and angular velocity inputs, respectively. The measurement $y \in \R$ measures the strength of the field $F(x)$ at position $x$ of the vehicle. 
	The shape of the signal field $F(x)$ is generally unknown, however we make the following assumption on its form.
	\begin{assumption} 
		\label{ass:shape_of_field}
		Let $\kappa \in \mathbb{N}$. The function $F \in \mathcal{C}^2: \R^2 \rightarrow \R$ is radially unbounded and has an unknown global maximum, also referred to as 'source',  at position $x^*$, i.e., 
		$F(x^*) \geq F(x)$ for $\forall x \neq x^*$ with $\nabla F(x^*) = 0$. Moreover, $F(x)$ is bounded as 
		\begin{equation}
		F(x^*) - a_1 \norm{x-x^*}^{2\kappa} \leq F(x) \leq F(x^*) - a_2 \norm{x-x^*}^{2\kappa} 
		\label{eq:assumption2}
		\end{equation}
		where $a_1, a_2 \in (0, \infty)$, 
		and $\nabla F(x) \in \R^{2 \times 1}$ is bounded as
		\begin{equation}
		b_1 \norm{x - x^*}^{2 \kappa -1} \leq \norm{\nabla F(x) } \leq b_2 \norm{x - x^*}^{2 \kappa -1}
		\label{eq:assumption3}
		\end{equation}
		where $b_1, b_2 \in (0, \infty)$.
		
	\end{assumption}
The reason for Assumption~\ref{ass:shape_of_field} will become apparent in the later sections. Furthermore, we view the unknown drift term $f(t,x)$ as a perturbation in the nominal (unicycle) system \eqref{eq:unicycle_with_drift-a}. We distinguish two cases, covered by the following assumption.
	\begin{assumption}
		\label{ass:drift_term}
		The perturbation $f \in \C^2: [t_0, t_0 + T) \times \R^2 \rightarrow \R^2$  in \eqref{eq:unicycle_with_drift-a} is 
		\begin{enumerate}[label=\Alph*.,ref=\ref{ass:drift_term}.\Alph*]
			\item either vanishing, i.e., $f(t,x^*) = 0$, and additionally $f(t,x) = E \nabla F(x)$ with an arbitrary unknown $E \in \R^{2 \times 2}$, \label{ass:drift_A}
			\item or nonvanishing with  
    $\norm{f(t,x)} \le d(t)$, where 
    $\norm{d}_{[t_0, t]} := \sup\limits_{t_0 \le s \le t} \abs{d(s)} $ is uniformly bounded on $[t_0, t_0+T]$.  \label{ass:drift_B}
		\end{enumerate}
	\end{assumption} 
With the first case covered by Assumption \ref{ass:drift_A} where $f(t,x)$ is a gradient-dependent vanishing drift $E \nabla F(x)$, we model a repulsive source. An example of such a repellent source is found in the height-seeking application in \cite[Section~6.2]{10.1115/1.4025457} where the topography of the terrain, through gravity, alters the unicycle model of a ground vehicle and causes downslope slipping. In Assumption \ref{ass:drift_B}, we cover the case of a  nonvanishing drift term that has a more general but bounded form, which might result from disturbances or other physical causes.
The objective is to design a control law $u_1(t)$ and $u_2(t)$ for  \eqref{eq:unicycle_with_drift}, so that the position of the vehicle $x$ converges to the source $x^*$ of the field $F(x)$ in prescribed finite time $T$, i.e., as $t \rightarrow t_0 + T$ without using the position information $x$ and despite the influence of the possibly unstable drift term $f(t,x)$.
	

	\section{PT Source Seeking Design}\label{sec:design}
	Following the methodology presented in Section 2.2, we replace the control objective from Section \ref{sec:problem_statemenet} by examining  \eqref{eq:unicycle_with_drift} in time $\tau$. For this, we perform the time dilation $t \rightarrow \tau$ on \eqref{eq:unicycle_with_drift} with \eqref{eq:dt_dtau} and obtain 
	\begin{subequations}
		\label{eq:seeker_tau_to_be_designed}
		\begin{align}
		\frac{\text{d} x}{\text{d} \tau}   &=\frac{f(\tau,x)}{\left(1 + \frac{\tau-t_0}{T} \right)^2}  + \frac{1}{\left(1 + \frac{\tau-t_0}{T} \right)^2} u_1(\tau) \begin{bmatrix}
		\cos(\theta) \\
		\sin(\theta)
		\end{bmatrix} \\
		\frac{\text{d} \theta}{\text{d} \tau}    &=  \frac{1}{\left(1 + \frac{\tau-t_0}{T} \right)^2} u_2(\tau).
		\end{align}
	\end{subequations}
	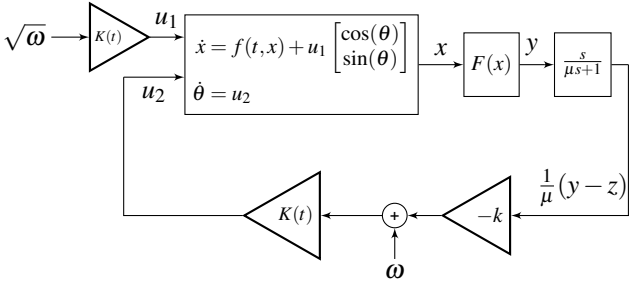
\begin{figure}[t]
		\centering
		\vspace{0.15cm}
		\begin{tikzpicture}[auto, node distance=2cm,>=latex',scale = 0.6]
		
		\node [block, scale = 0.8] (system) {$\begin{aligned}
			\dot{x} &= f(t,x) + u_1 \begin{bmatrix}
			\cos(\theta) \\
			\sin(\theta)
			\end{bmatrix} \\
			\dot{\theta} &= u_2
			\end{aligned}$};
		
		\node[gain, above left= -0.5cm and 0.8 cm of system, node distance = 0mm, scale = 0.7] (in)   {$K(t)$};
		
		\node[input, below left= -0.5 cm and 0.8cm of system] (help)  {};
		
		\node[input, left = 0.5cm of in] (leftin) {};
		
		\node [block, right of=system, node distance=2.5cm, scale = 0.8] (Fx) {$F(x)$};
		\draw [->] (system) -- node[name=u] {$x$} (Fx);
		
		\node [output, right of=Fx, node distance = 0.7cm] (output) {};
		\node [gain, below = 2cm of u, shape border rotate= 180, anchor = apex] (gain) {$-k$};
		\node [sum, left of=gain, node distance=2.5cm] (sum1) {\large \textbf{+}};
		\node [below of=sum1, node distance = 0.7cm] (omega) {};
		\draw [->] (help) -- node[anchor = north] {$u_2$} (system.west |- help);

		\draw [->] (leftin) -- node[anchor = east, pos = 0.1] {$\sqrt{\omega}$}  (in);
		\draw [->] (in) -- node[midway] {$u_1$}  (system.west |- in);
		\draw [] (Fx) -- node {$y$} (output);
		
		\node[block, right of = output, node distance  = 0.5cm, scale = 0.8] (filter) {$\frac{s}{\mu s + 1}$};
		\node [output, right of=filter, node distance = 0.6cm] (outputfilter) {};
		
		\node [gain, left of=sum1, shape border rotate= 180, anchor = apex, node distance = 2.5cm,, scale = 0.9] (gaint) {$K(t)$};
		
		\draw[->] (sum1) -- (gaint);
		\draw [-] (help) |- (gaint);
		\draw [->] (output) -- (filter);
		\draw [-] (filter) -- (outputfilter);
		\draw [->] (outputfilter) |-node[anchor = south, xshift = -0.6cm] {$\frac{1}{\mu}(y-z)$} (gain);
		\draw [->] (gain) -- (sum1);
		\draw [->] (omega) -- node[anchor = north, pos = 0.1] {$\omega$} (sum1);
		
		\end{tikzpicture}
		\caption{PT Seeker for the nonholonomic unicycle with a possibly unstable perturbation $f(t,x)$ where $K(t) = 1 / \nu^2(t-t_0)$.} 	
		\label{fig:unicycle_pt_controller}
	\end{figure}
	Thus, the new control objective in time $\tau$ becomes reaching the repulsive source $x^*$ as $\tau \rightarrow \infty$. 
	Motivated by the asymptotic seeker proposed in \cite[Section 4]{ES_using_LBAandSP}, for which the states $x$ are proven to converge to $x^*$ as $\tau \rightarrow \infty$, we choose the time-varying dynamic feedback law as
	\begin{subequations}
		\label{eq:seeker_tau}
		\begin{align}
		u_1 (\tau) &= \left(1 + \frac{\tau - t_0}{T}\right)^2 \sqrt{\omega} \\
		u_2(\tau) &= \left(1 + \frac{\tau - t_0}{T}\right)^2  \left( \omega - \frac{k}{\mu} (y-z)\right)\\
		\frac{{\rm d} z}{{\rm d} \tau} &=  \frac{1}{\mu} (y-z) \label{eq:filter_tau}.
		\end{align}
	\end{subequations}
	The equation \eqref{eq:filter_tau} comes from the fact that the time derivative of the measurement $y(t)$  is replaced by a washout filter $\frac{s}{\mu s + 1}$, namely, by an approximate differentiator. The signal after the washout filter is then $\frac{1}{\mu} (y-z)$, where $z$ denotes the state of the filter. Substituting \eqref{eq:seeker_tau} in \eqref{eq:seeker_tau_to_be_designed} yields
	\begin{subequations}
		\label{eq:closed_loop_system_tau}
		\begin{align}
		\frac{\text{d} x }{\text{d} \tau}  &= 
		\frac{f(\tau,x)}{(1+ \frac{\tau-t_0}{T})^2}  + \sqrt{\omega} \begin{bmatrix}
		\cos(\theta) \\
		\sin(\theta)
		\end{bmatrix}
		\\
		\frac{\text{d} \theta }{\text{d} \tau}  &= \omega - \frac{k}{\mu} (y-z)  \\
		\frac{\text{d} z }{\text{d} \tau}  &=  \frac{1}{\mu}(y-z). 
		\end{align}
	\end{subequations}
      The mechanism behind the convergence of the system \eqref{eq:closed_loop_system_tau} to the source $x^*$ is found in \cite[Sec.~1]{ES_with_bounded_update_rates}. Notice that, due to the drift term $f(\tau,x)/(1+ \frac{\tau-t_0}{T})^2 $, the proposed closed-loop system for source seeking \eqref{eq:closed_loop_system_tau} is  different from the one shown in \cite{ES_using_LBAandSP} and the stability of the system needs to be analyzed separately.
	Next, we perform the time contraction $\tau \rightarrow t$ with  \eqref{eq:dtau_dt} and obtain \eqref{eq:closed_loop_system_tau} in time $t$ as
	\begin{subequations}
		\label{eq:closed_loop_system_t}
		\begin{align}
		\frac{{\rm d} x}{{\rm d} t} &= f(t,x) + \frac{1  }{v^2(t-t_0)} \sqrt{\omega} \begin{bmatrix}
		\cos(\theta) \\
		\sin(\theta)
		\end{bmatrix} \\
		\frac{{\rm d} \theta}{{\rm d} t} &= \frac{1}{v^2(t-t_0)}\left(\omega - \frac{k}{\mu} (y-z) \right) \\
		\frac{{\rm d} z}{{\rm d} t} &= \frac{1}{v^2(t-t_0)} \frac{1}{\mu}(y-z).
		\end{align}
	\end{subequations}
	The closed-loop system \eqref{eq:closed_loop_system_t} is referred to as PT Seeker (see Fig.~\ref{fig:unicycle_pt_controller}). From it,  we deduce the PT feedback law given in the next section. The corresponding stability properties are defined in the Appendix. 
	
	\section{PT Convergence Analysis}
	\label{sec:convergence_analysis}
		\begin{theorem}[PT seeking under {\em vanishing} drift] \label{thm:main_theorem}

			Suppose that the map $F$ satisfies Assumption \ref{ass:shape_of_field} and consider the unicycle \eqref{eq:unicycle_with_drift} with the  feedback 
			\begin{subequations}
				\label{eq:pt_controlla}
				\begin{align}
				u_1(t) &= \frac{1 }{v^2(t-t_0)}\sqrt{\omega} \label{eq:u1}\\
				u_2(t) &= \frac{1 }{v^2(t-t_0)}\left(\omega - \frac{k}{\mu} (y-z) \right) \label{eq:u2} \\
				\dot{z} &= \frac{1 }{v^2(t-t_0)}\frac{1}{\mu}(y-z) \label{eq:z}
				\end{align}
			\end{subequations}
			where  $\nu(t-t_0)$ is defined in \eqref{eq:nu}.
   The source $x=x^*\in\R^2$ is FxT-sSPUAS (with the roles of $T,\omega,\mu$ attributed as in Def. \ref{def:FxT-sSPUAS})
   in either of the following cases:
			\begin{enumerate}[label=\roman*.,ref=\ref{thm:main_theorem}.\roman*]
				\item  $f(t,x) \equiv 0\in\R^2$, \label{it:theorem_part_1}  
				\item  $f(t,x)$ satisfies Assumption \ref{ass:drift_A} and  $\kappa = 1$. \label{it:theorem_part_2}
			\end{enumerate}
		\end{theorem}
\begin{proposition}[PT seeking under {\em non-vanishing} drift]
\label{proposition}
    Consider the Lie bracket averaged reduced model of \eqref{eq:closed_loop_system_t}, given by  
    \begin{equation}
      \frac{\text{d} \bar{x}}{\text{d} t} = f(t,\bar{x}) + \frac{1}{v^2(t-t_0)} \frac{k}{2}\nabla F(\bar{x}). \label{eq:LBA_reduced_model_t}
    \end{equation}
    Under Assumptions \ref{ass:shape_of_field} and \ref{ass:drift_B}, the state $\bar{x}(t)$ of \eqref{eq:LBA_reduced_model_t} converges to the source $x^*$  as $t\rightarrow t_0 + T $.
\end{proposition} 

	\noindent{\bf Proof of Theorem \ref{thm:main_theorem}.}
	The proof is divided in 6 steps.
	
	\vspace*{.5em}\noindent{ \textbf{Step 1.} \underline{\textit{Time Dilation $t \rightarrow \tau$}:}}
	In this step we transform \eqref{eq:closed_loop_system_t} back in time $\tau$ with \eqref{eq:dt_dtau} and obtain \eqref{eq:closed_loop_system_tau}.   
	Next, we introduce the change of variables $p = \theta + kz$, where $\frac{\text{d} p}{ \text{d} \tau } = \frac{\text{d} \theta}{ \text{d} \tau} + k\frac{\text{d} z}{ \text{d} \tau} = \omega$ and $p(\tau) = \omega \tau + p_0 $. Without loss of generality, we set the initial orientation $p_0 = 0$. Substituting $\theta = p - kz $ in \eqref{eq:closed_loop_system_tau}, yields 
	
	\begin{subequations}
		\label{eq:tau_pt_system}
		\begin{align}
		\frac{\text{d} x}{\text{d}\tau } &=  
		\frac{f(\tau,x)}{(1+ \frac{\tau-t_0}{T})^2}   +  \sqrt{\omega} \begin{bmatrix}
		\cos(\omega \tau - kz ) \\
		\sin(\omega  \tau - kz )
		\end{bmatrix}
		\\
		\frac{\text{d} z}{\text{d}\tau } &=  \frac{1}{\mu} (y - z) 
		\end{align}
	\end{subequations}
	\vspace*{.5em}\noindent{\textbf{Step 2.} \underline{\textit{Singular Perturbation}:}}
	In order to use the analysis introduced in Section 2.1, consider the time scale change for \eqref{eq:tau_pt_system} as $\tilde{\tau} = \frac{\tau}{\mu}$ and $\frac{\text{d} \tau}{\text{d} \tilde{\tau}} = \mu$ which yields
	\begin{subequations}
		\label{eq:tau_pt_system_eps}
		\begin{align}
		\frac{\text{d} x}{\text{d}\tilde{\tau} } &= \begin{aligned}[t]
		\frac{ \mu f(\tilde{\tau},x)  }{(1+ \frac{\mu\tilde{\tau}-t_0}{T})^2} + \mu \sqrt{\omega} \begin{bmatrix}
		\cos(\omega \mu \tilde{\tau} - kz ) \\
		\sin(\omega \mu \tilde{\tau} - kz )
		\end{bmatrix}
		\end{aligned} \label{eq:to_be_reduced_model} \\
		\frac{\text{d} z}{\text{d} \tilde{\tau} } &=  y - z.
		\end{align}
	\end{subequations}
	Since \eqref{eq:tau_pt_system_eps} is of the form  \eqref{eq:general_nominal_system}, we can apply the singularly pertubed Lie bracket approximation analysis.
	Recognizing that $l(x) = y = F(x)$, the \textit{boundary layer} model for \eqref{eq:tau_pt_system_eps} is obtained as
	\begin{equation}
	\label{eq:blm_real}
	\frac{\text{d} \tilde{y}}{\text{d} \tilde{\tau} } = g(x, \tilde{y}+l(x))= - \tilde{y}. 
	\end{equation}
	The \textit{reduced} model is derived by substituting the quasi steady-state $z = l(x) =F(x)$ in \eqref{eq:to_be_reduced_model} which yields
	\begin{equation}
	\frac{\text{d} \tilde{x}}{\text{d} \tilde{\tau} } = 
\frac{ \mu f(\tilde{\tau},\tilde{x})  }{(1+ \frac{\mu\tilde{\tau}-t_0}{T})^2} + \mu \sqrt{\omega} \begin{bmatrix}
	\cos(\omega \mu \tilde{\tau} - kF(\tilde{x}) ) \\
	\sin(\omega \mu \tilde{\tau} - kF(\tilde{x}) )
	\end{bmatrix} \label{eq:reduced_model_tau}
	\end{equation}
	\vspace*{.5em}\noindent{\textbf{Step 3.} \underline{\textit{Lie Bracket Averaging}:}} Now, we derive the Lie bracket system for the \textit{reduced} model \eqref{eq:reduced_model_tau}. Using the trigonometric identities $\cos(\alpha + \beta) = \cos(\alpha) \cos(\beta) - \sin(\alpha) \sin(\beta) $ and $\sin(\alpha + \beta) = \sin(\alpha) \cos(\beta) + \cos(\alpha) \sin(\beta)$, \eqref{eq:reduced_model_tau} becomes 
	{\setlength{\mathindent}{0pt}
		\begin{align}
		&\frac{\text{d} \tilde{x}}{\text{d} \tilde{\tau} } = 
		\underbrace{\frac{ \mu f(\tilde{\tau},\tilde{x})  }{(1+ \frac{\mu\tilde{\tau}-t_0}{T})^2}}_{b_0(\mu \tilde{\tau},\tilde{x}, l(\tilde{x}))/ \mu}  + \mu \sqrt{\omega}  \underbrace{ \begin{bmatrix} 
			\cos(kF(\tilde{x}) ) \\
			-\sin( kF(\tilde{x}))
			\end{bmatrix}}_{b_1(\mu \tilde{\tau},\tilde{x},l(\tilde{x}))} \underbrace{\cos(\omega \mu \tilde{\tau})}_{u_1(\omega \mu \tilde{\tau})} \nonumber  \\ &+  \mu \sqrt{\omega} \underbrace{\begin{bmatrix}
			\sin(kF(\tilde{x}) ) \\
			\cos( kF(\tilde{x}))
			\end{bmatrix}}_{b_2(\mu \tilde{\tau},\tilde{x},l(\tilde{x}))} \underbrace{\sin(\omega \mu \tilde{\tau} )}_{u_2(\omega \mu \tilde{\tau})} 
		\label{eq:reduced_model_tau_2}
		\end{align}
	}
	Defining  $b_i(\mu \tilde{\tau}, \tilde{x})$, $i = 0,1,2$ and $u_j(\omega \mu \tilde{\tau})$, $j = 1,2$ as in \eqref{eq:reduced_model_tau_2} and  $\mu = 1$ which implies $\tilde{\tau} = \tau$, we write the Lie bracket system of \eqref{eq:reduced_model_tau_2} similar as in \cite{ES_using_LBAandSP} as
	\begin{align}
	\frac{\text{d} \bar{x}}{\text{d} \tau} = \frac{f(\tau,\bar{x})}{(1+\frac{\tau-t_0}{T})^2} + \frac{k}{2} \nabla F(\bar{x}). \label{eq:lie_bracket_system_real}
	\end{align}
	Since $\mu$ plays the role of a time
	scale, setting $\mu = 1$ does not influence the qualitative stability properties  (see \cite{Singularly_Pertubed_LBA}).

	\vspace*{.5em}\noindent{ \textbf{Step 4.}  \underline{\textit{Stability Analysis of the Lie Bracket System}:} }
	In what follows, we analyze the stability of the Lie bracket system \eqref{eq:lie_bracket_system_real}.
	For this, we choose the Lyapunov function candidate
	\begin{align}
	V(\bar{x}) = - (F(\bar{x}) - F(x^*)) \label{eq:Lyapunov_function}
	\end{align}
	whose derivative along the solutions of \eqref{eq:lie_bracket_system_real} is
	\begin{align}
	\frac{\text{d} V(\bar{x})}{\text{d} \tau} &= - \frac{1}{(1+\frac{\tau-t_0}{T})^2}  \nabla F(\bar{x})^\top f(\tau,\bar{x}) -  \frac{k}{2}  \norm{\nabla F(\bar{x})}^2. \label{eq:Lyapunov_derivative}
	\end{align}
	Using the inequality $-\nabla F(\bar{x})^\top f(\tau,\bar{x})  \le \ \norm{f(\tau,\bar{x})} \norm{\nabla F(\bar{x})}$, we obtain an upper bound of \eqref{eq:Lyapunov_derivative} as
	\begin{align}
	\frac{\text{d} V(\bar{x})}{\text{d} \tau} \le \frac{1}{(1+\frac{\tau-t_0}{T})^2} \norm{f(\tau,\bar{x})} \norm{\nabla F(\bar{x})}- \frac{k}{2} \norm{\nabla F(\bar{x})}^2 . \label{eq:Lyapunov_derivative_inequality}
	\end{align}
	Rearranging \eqref{eq:assumption2} and taking \eqref{eq:Lyapunov_function} into account , we can write
	\begin{align}
	a_2 \norm{\bar{x} - x^*}^{2\kappa} &\leq V(\bar{x}) \leq a_1 \norm{\bar{x} - x^*}^{2\kappa}. \label{eq:ass2} 
	\end{align}
	Combining  \eqref{eq:assumption3} and  \eqref{eq:ass2}, we bound $\norm{\nabla F(x)}$ as
	\begin{align}
	 c_{\kappa,1} V(\bar{x})^{q_{\kappa}} \leq \norm{\nabla F(\bar{x})} \leq c_{\kappa,2} V(\bar{x})^{q_{\kappa}}. \label{eq:ass4}
	\end{align}
	where $q_{\kappa} = \frac{2 \kappa - 1}{2 \kappa} \in [1/2, 1)$ and $c_{\kappa,i} = \frac{b_i}{a_i^{q_{\kappa}}} > 0$ with $i = 1,2$. Thus, with \eqref{eq:ass4}, the upper bound \eqref{eq:Lyapunov_derivative_inequality} becomes
\begin{equation}
		\hspace*{-0.15cm}
		\frac{\text{d} V(\bar{x})}{\text{d} \tau} \le \frac{c_{\kappa,2}}{(1+\frac{\tau-t_0}{T})^2}  \norm{f(\tau,\bar{x})} V(\bar{x})^{q_{\kappa}}- \frac{k c^2_{\kappa,1}  }{2} V(\bar{x})^{2 q_{\kappa}}. \label{eq:lyapunov_derivative_inequality_2}
\end{equation}
	We now consider the two cases from Theorem \ref{thm:main_theorem}:
	\begin{enumerate}[label = \roman*.]
		\item Since $f(t,x) \equiv 0$, then $\norm{f(\tau,x)} =0$. Thus, \eqref{eq:lyapunov_derivative_inequality_2} becomes 
		\begin{equation}
		\frac{\text{d} V(\bar{x})}{\text{d} \tau} \le   - \frac{k c^2_{\kappa,1}  }{2} V(\bar{x})^{2 q_{\kappa}} \label{eq:lyapunov_derivative_inequality_kappa_2}.
		\end{equation}
		Considering Assumption \ref{ass:shape_of_field} and $k c^2_{\kappa,1} > 0$, we  conclude that \eqref{eq:lyapunov_derivative_inequality_kappa_2} satisfies the conditions from \cite[Theorem 4.9]{khalil2002nonlinear}. This implies that the equilibrium point $x^*$ of   \eqref{eq:lie_bracket_system_real} is GUAS in time $\tau \in I_{\tau}$.
		
		\item Due to Assumption \ref{ass:drift_A} and $\kappa = 1$, and with \eqref{eq:ass4}, $\norm{E \nabla F(x)} \le \norm{E} \norm{\nabla F(x)}$ and $2 q_{1} = 1$, the differential inequality \eqref{eq:lyapunov_derivative_inequality_2} becomes
		\begin{equation}
		\frac{\text{d} V(\bar{x})}{\text{d} \tau} \le \frac{1}{(1+\frac{\tau-t_0}{T})^2} c^2_{1,2} \norm{E}  V(\bar{x}) -    \frac{k c^2_{1,1}}{2} V(\bar{x}). \label{eq:lyapunov_derivative_inequality_kappa_1}
		\end{equation}
		Furthermore, by performing the substitution $\hat{\tau} = \tau - t_0$ on \eqref{eq:lyapunov_derivative_inequality_kappa_1}, we can use \cite[Lemma B.3]{krstic2010delay} with $\norm{l_1}_1 = c^2_{1,2} \norm{E} T $, $l_2(\tau) = 0$ and $c = k c^2_{1,1} / 2$ and obtain the upper bound of $V(\bar{x})$ for $\kappa = 1$ in the original time $\tau$ as
		\begin{equation}
		V(\bar{x}) \le V(\bar{x}(t_0)) e^{c^2_{1,2} \norm{E} T} e^{- \frac{k c^2_{1,1}}{2} (\tau - t_0) }. \label{eq:Lyapunov_inequality_upper_bound}
		\end{equation}
		Considering \eqref{eq:ass2} and \eqref{eq:Lyapunov_inequality_upper_bound}, we get an upper bound for the norm of $\bar{x}$ as
		\begin{equation}
		\hspace*{-0.5cm}
		\norm{\bar{x}(\tau) - x^*} \le \sqrt{\frac{a_1}{a_2}} \norm{\bar{x}(t_0) - x^*} e^{ \frac{c^2_{1,2}}{2} \norm{E} T} e^{- \frac{k c^2_{1,1}}{4} (\tau-t_0) } \label{eq:x_bar_upper_bound}
		\end{equation}
		From \eqref{eq:x_bar_upper_bound}, it is obvious that as $\tau \rightarrow \infty$, the state $\bar{x} \rightarrow x^*$ which implies that the equilibrium point $x^*$ of \eqref{eq:lie_bracket_system_real} is GUAS in time $\tau \in I_{\tau}$.
	\end{enumerate} 
\noindent{\textbf{Step 5.} \underline{\textit{Singularly perturbed Lie Bracket Averaging Theorem}:}}
	In Step 4,  we showed the asymptotic stability properties of the Lie bracket system \eqref{eq:lie_bracket_system_real}. 
	In addition, we observe that by choosing $V(\tilde{y}) = \frac{1}{2}{\tilde{y}}^2$ with $\frac{\partial V(\tilde{y} )}{\partial \tilde{y}} g(x, \tilde{y}+l(x)) = -\tilde{y}^2$ the \textit{boundary layer} model \eqref{eq:blm_real} from Step 2 satisfies the requirements \eqref{eq:blm_req_1} and \eqref{eq:blm_req_2}. Hence, we apply Theorem \ref{thm:SPLBave} from Section 2.1 and conclude that the equilibrium point, i.e., the source $x^*$ of \eqref{eq:closed_loop_system_tau} is sSPUAS on the infinite time interval $I_{\tau}$ in both cases of Theorem \ref{thm:main_theorem}. 

\vspace*{.5em}\noindent{\textbf{Step 6.} \underline{\textit{Time Contraction $\tau \rightarrow t$}:}} 
	In the final step, we convert \eqref{eq:closed_loop_system_tau} to \eqref{eq:closed_loop_system_t} using \eqref{eq:dtau_dt}.
	The temporal transformation \eqref{eq:t_to_tau} retains the stability properties of \eqref{eq:closed_loop_system_tau} in time $t$, meeting, in particular,  Definition~\ref{def:FxT-sPUA}. This implies that the equilibrium point $x^*$ of \eqref{eq:closed_loop_system_t} is sSPUAS on the finite time interval $I_t$ (FxT-sSPUAS) which immediately leads to the statement of Theorem~\ref{thm:main_theorem}. 
	\hfill Q.E.D.
		
	\vspace*{.5em} \noindent{\bf Proof of Proposition \ref{proposition}.} The Lie bracket averaged reduced model \eqref{eq:LBA_reduced_model_t} is obtained by transforming \eqref{eq:lie_bracket_system_real} back in time $t$ with \eqref{eq:t_to_tau}. The rest of Proposition \ref{proposition} is derived by examining \eqref{eq:lie_bracket_system_real} in time $\tau$. We continue our analysis as in Step 4 of the proof of Theorem \ref{thm:main_theorem} and obtain \eqref{eq:lyapunov_derivative_inequality_2}, which 
	under Assumption~\ref{ass:drift_B},  with $\norm{f(\tau,x)} \le d(\tau) \le \norm{d}_{[t_0,\tau]}$, becomes 
	\begin{equation}
	\hspace*{-0.5cm}
	\frac{\text{d} V(\bar{x})}{\text{d} \tau} \le  \frac{c_{\kappa,2} \norm{d}_{[t_0,\tau]} }{(1+\frac{\tau-t_0}{T})^2} V(\bar{x})^{q_{\kappa}} - \frac{k c^2_{\kappa,1}  }{2} V(\bar{x})^{2 q_{\kappa}}. \label{eq:lyapunov_derivative_inequality_bounded}
	\end{equation}
	Substituting $U(\bar{x}) = V(\bar{x})^{1-q_{\kappa}} = V(\bar{x})^{1/2\kappa}$ in \eqref{eq:lyapunov_derivative_inequality_bounded} with ${\rm d} U / {\rm d} V = (1-q_{\kappa}) V^{-q_{\kappa}}$  yields
	\begin{equation}
	\frac{\text{d} U(\bar{x})}{\text{d} \tau} \le -h_1 U(\bar{x})^{2\kappa - 1} + \underbrace{ h_2  \norm{d}_{[t_0,\tau] } / (1+(\tau-t_0)/T)^2 }_{\zeta(\tau)} \label{eq:Ux_inequality}
	\end{equation} 
	where  $h_1 =   k c^2_{\kappa,1} / 4\kappa$ and $h_2 = c_{\kappa,2}  / 2 \kappa$. Since $V(\bar{x}) = U(\bar{x})^{2 \kappa}$, from \eqref{eq:ass2}, we  write 
	\begin{equation}
	a_2^{1/2\kappa} \norm{\bar{x} - x^*} \leq U(\bar{x}) \leq a_1^{1/2\kappa} \norm{\bar{x} - x^*} \label{eq:Ux_bounds}
	\end{equation}
	With \eqref{eq:Ux_inequality} and \eqref{eq:Ux_bounds}, we derive an upper bound for $\frac{\text{d} U(\bar{x})}{\text{d} \tau}$ as 
	\begin{equation}
	\frac{\text{d} U(\bar{x})}{\text{d} \tau} \le -h_1 a_2^{(2\kappa-1)/ 2\kappa} \norm{\bar{x}-x^*}^{2\kappa - 1} + \norm{\zeta(\tau)} \label{eq:Ux_inequality2}
	\end{equation}
	From \eqref{eq:Ux_inequality2}, it follows that 
	\begin{equation}
	\hspace*{-0.3cm}
	\begin{aligned}
	\frac{\text{d} U(\bar{x})}{\text{d} \tau} \le -h_1 a_2^{\frac{r}{2\kappa}} \norm{\bar{x}-x^*}^{r}, \quad  \forall \norm{\bar{x} - x^*} \ge  \sqrt[r]{\frac{\norm{\zeta(\tau)}}{a_2^{\frac{r}{2\kappa}}h_1}} 
	\end{aligned} \label{eq:Ux_inequality3}
	\end{equation} 
	with $r =  2\kappa -1 $. From \eqref{eq:Ux_bounds} and \eqref{eq:Ux_inequality3}, it is clear that we can apply \cite[Theorem 4.19]{khalil2002nonlinear} and conclude that \eqref{eq:lie_bracket_system_real} is \textit{input-to-state} stable (ISS)  w.r.t. $\zeta(\tau)$ with $\gamma(\tilde{s}) = (a_1 / a_2^2)^{1/2\kappa} \sqrt[r]{\tilde{s}/h_1}$. 
	By  \cite[Def. 4.7]{khalil2002nonlinear},  ISS  implies  that 
	\begin{equation}
	\hspace*{-0.5cm}
	\norm{\bar{x}(\tau) - x^*} \le \beta(\norm{\bar{x}(p) - x^*}, \tau - p) + \gamma\left( \norm{\zeta}_{[p,\tau]} \right) \label{eq:upper_bound}
	\end{equation}
	where $\zeta(\tau)$ is defined as in \eqref{eq:Ux_inequality}, $\norm{\zeta}_{[p,\tau]} = \sup\limits_{p\le s \le \tau} \abs{\zeta(s)}$, $\tau>p\ge t_0$, $\beta$ is a class $\mathcal{K} \mathcal{L}$ function and $\gamma$ is class $\mathcal{K}$ defined as above.	
	Substituting $(p,\tau) = (t_0, (t_0 + \tau)/2)$ in \eqref{eq:upper_bound}, yields
	\begin{equation}
	\hspace*{-0.3cm}
	\begin{aligned}
	\norm{\bar{x}((t_0 + \tau)/2)  - x^*} \le &\beta\left(\norm{\bar{x}(t_0) - x^*}, (\tau - t_0)/2\right) + \\ &\gamma\left( h_2 \norm{d}_{\left[t_0, \tau \right] }    \right). 
	\end{aligned} \label{eq:upper_bound1}
	\end{equation}
	Then, with $(p,\tau) = ((t_0 + \tau)/2,\tau)$ in \eqref{eq:upper_bound}, we write
	\begin{equation}
	\hspace*{-0.5cm}
	\begin{aligned}
	\norm{\bar{x}(\tau) - x^*} &\le \beta(\norm{\bar{x}((t_0 + \tau )/2) - x^*}, (\tau-t_0)/2) +\\ &\gamma\left( h_2 \norm{d}_{\left[t_0, \tau \right] } / \left( 1 + (\tau - t_0) / 2T\right)^2 \right).
	\end{aligned}
	\label{eq:upper_bound2}
	\end{equation}
	Combining \eqref{eq:upper_bound1} and  \eqref{eq:upper_bound2}, we obtain
	\begin{equation}
	\hspace*{-0.5cm}
	\begin{split}
	&\norm{\bar{x}(\tau) - x^*} \le \beta\left(\beta\left(\norm{\bar{x}(t_0) - x^*}, \frac{\tau - t_0}{2}\right) \right.  \\  &+ \gamma \left( h_2 \norm{d}_{\left[t_0, \tau \right]}\right) ,  \left. \frac{\tau - t_0}{2} \right) + 
	\gamma\left( \frac{h_2  \norm{d}_{\left[t_0, \tau \right] } } {\left( 1 + \frac{\tau - t_0}{2T}  \right)^2} \right)		\end{split}
	\label{eq:finale}
	\end{equation}
	Since $\lim\limits_{\tau \rightarrow \infty}\frac{h_2  \norm{d}_{\left[t_0, \tau \right] } }{\left( 1 + \frac{\tau - t_0}{2T}  \right)^2} = 0$ in \eqref{eq:finale}, we can infer that $\bar{x} \rightarrow x^*$ as $\tau \rightarrow \infty$. 
	Then, transforming \eqref{eq:finale} back in time $t$ with \eqref{eq:t_to_tau} yields 
	\begin{equation}
		\begin{split}
			&\norm{\bar{x}(t) - x^*} \le \beta\left(\beta\left(\norm{\bar{x}(t_0) - x^*}, \frac{t - t_0}{2 \nu(t-t_0)}\right) \right.  \\  &+ \gamma\left( h_2 \norm{d}_{\left[t_0, t \right] }   \right),  \left. \frac{t - t_0}{2 \nu(t-t_0)} \right) \\ &+ 
	\gamma\left( \frac{h_2  \norm{d}_{\left[t_0, t \right] } }{\left( 1 + \frac{t-t_0}{2T \nu(t-t_0)}  \right)^2} \right).	\end{split}
		\label{eq:finale_t}
	\end{equation}
	Letting $t \rightarrow t_0 + T$ in \eqref{eq:finale_t}, immediately leads to the statement in Proposition \ref{proposition}.
\hfill Q.E.D.

\section{Discussion on PT Source Seeking } 
	\label{sec:feasibility}
	In this section, we discuss the consequences of Theorem \ref{thm:main_theorem} and Proposition \ref{proposition} and emphasize the advantages of the PT Seeker \eqref{eq:closed_loop_system_t}. For this, we compare \eqref{eq:closed_loop_system_t} with its asymptotic counterpart. We get the asymptotic Seeker from \eqref{eq:closed_loop_system_t} by considering $\lim\limits_{T \rightarrow \infty} \frac{1}{\nu^2(t-t_0)} = 1$ as
	\begin{subequations}
		\label{eq:exponential_seeker}
		\begin{align}
		\dot{x} &= f(t,x) +  \sqrt{\omega} \begin{bmatrix}
		\cos(\theta) \\
		\sin(\theta)
		\end{bmatrix} \\
		\dot{\theta} &= \omega - \frac{k}{\mu} (y-z)  \\
		\dot{z} &=  \frac{1}{\mu}(y-z).
		\end{align}
	\end{subequations}
It is important to note that, while the conditions of Assumption \ref{ass:shape_of_field} are rarely satisfied globally in physical problems (usually, signals decay to zero away from the source), they are instrumental in getting semiglobal stability results. If the signal decays to zero far from the source, a start from a sufficiently remote location, where the gradient is sufficiently low, results in the vehicle barely progressing to the source, apperaing that it is stuck, in spite of the average system \eqref{eq:lie_bracket_system_real} suggesting attractivity, since the {\em globality} assumption of the Lie bracket averaging Theorem \ref{thm:SPLBave} is not met due to the absence of radial unboundedness of the Lyapunov function defined through the map $F$, \eqref{eq:Lyapunov_function}. Note that, if Assumption \ref{ass:shape_of_field} does not hold globally, the results are local.

\textit{6.1.~Driftless case}: 
In the absence of a drift term, i.e., $f(t,x) =0$, the asymptotic Seeker \eqref{eq:exponential_seeker} on average converges to the source as $t \rightarrow \infty$, as shown in \cite{ES_using_LBAandSP}, whereas the PT Seeker \eqref{eq:closed_loop_system_t} reaches the source $x^*$ in user-assignable prescribed time $T$ as $t \rightarrow t_0 + T$, as stated in Theorem \ref{it:theorem_part_1}. PT convergence is obviously advantageous for time-critical applications.

\textit{6.2. Vanishing drift term:}	Let us now focus on the case where $f(t,x) = E \nabla F(x)$ and the map $F(x)$ is quadratically  bounded, i.e.,  $\kappa = 1$. The convergence analysis for the asymptotic Seeker \eqref{eq:exponential_seeker} is done similarly as in Section \ref{sec:convergence_analysis}. In order to deduce the stability properties of $x^*$ for \eqref{eq:exponential_seeker}, we make use of \eqref{eq:lyapunov_derivative_inequality_kappa_1} which becomes $\dot{V}(\bar{x}) \le  \left( c^2_{1,2} \norm{E}  -  k c^2_{1,1} / 2 \right) V(\bar{x})$.
From the negative definiteness condition on $\dot{V}(\bar{x})$ and by invoking Theorem \ref{thm:SPLBave}, we conclude that the asymptotic Seeker converges to $x^*$ as $t \rightarrow \infty$, provided 
\begin{equation}
k > 2 \left(c_{1,2} / c_{1,1} \right)^2 \norm{E}, \label{eq:convergence_cond}
\end{equation}
which is positive but unknown. In contrast to this, Theorem \ref{it:theorem_part_2} implies that starting from an arbitrarily large set of initial conditions, the trajectory of the PT Seeker \eqref{eq:closed_loop_system_t} converges to an unknown, but bounded neighborhood around the source $x^*$ in prescribed time $T$. This is further illustrated by inspecting \eqref{eq:x_bar_upper_bound} in time $t$, which in view of the time contraction \eqref{eq:t_to_tau} reads as
		\begin{equation}
		\hspace*{-0.4cm} \norm{\bar{x}(t) - x^*} \le \sqrt{\frac{a_1}{a_2}} \norm{\bar{x}(t_0) - x^*} e^{\frac{c_{1,2}^2}{2} \norm{E} T} e^{- \frac{k c^2_{1,1}}{4} \frac{t-t_0}{\nu(t-t_0)}} \label{eq:x_bar_upper_bound_in_t}.
		\end{equation}
		From  \eqref{eq:x_bar_upper_bound_in_t}, it is clear that on average $x \rightarrow x^*$ as $t \rightarrow t_0 + T$. Furthermore, we observe that the drift term $E\nabla F(x)$ reduces the speed of convergence, but does not influence the stability properties. Since the asymptotic Seeker has guaranteed convergence only when the gain $k$ satisfies \eqref{eq:convergence_cond}, which is not a priori verifiable, the PT Seeker \eqref{eq:closed_loop_system_t} offers a clear advantage.
 However, one limitation of PT source seeking becomes clear when inspecting Theorem \ref{it:theorem_part_2}, where it is obvious that the stability of $x^*$ can not be claimed when Assumption \ref{ass:shape_of_field} holds for $\kappa \ge 2$  and, in addition, Assumption \ref{ass:drift_A} is satisfied. The reason for this is as follows. Let $\kappa \ge 2$ and $f(t,x) = E\nabla F(x)$.
Substituting $\norm{E \nabla F(x) } \le c_{\kappa,2} \norm{E} V(\bar{x})^{q_{\kappa}} $ in \eqref{eq:lyapunov_derivative_inequality_2} and using the comparison principle, we obtain

\begin{equation}
\hspace*{-0.2cm}	V(\bar{x}(\tau)) \le V(x_0) \left( \frac{ \tau - t_0 + T }{\gamma_2 \tau^2 + \gamma_1 \tau + \gamma_0}\right)^{\frac{1}{2 q_{\kappa} - 1}} \label{eq:remark_inequality}
\end{equation}
where $\gamma_0 = (2 q_{\kappa} - 1) V(x_0)^{2 q_{\kappa} - 1} t_0 (c_{\kappa,2}^2 \norm{E} T +  0.5 k c^2_{\kappa,1}(t_0 -T)) - t_0 + T$, 	$\gamma_1 = 1 - (2 q_{\kappa} - 1) V(x_0)^{2 q_{\kappa} - 1} (c_{\kappa,2}^2 \norm{E} T - 0.5 k c^2_{\kappa,1}(T-2t_0) )$ and	$\gamma_2 = (2 q_{\kappa} - 1) V(x_0)^{2 q_{\kappa} - 1} 0.5 k c^2_{\kappa,1}$.
The denominator of \eqref{eq:remark_inequality} is a quadratic function of $\tau$ but, due to its linearity in $V(x_0)^{2q_{\kappa} - 1}$, we easily note that for any  $\tau^* > t_0$ and $\norm{E} > \frac{k c^2_{\kappa,1}}{2 c^2_{\kappa,2}} \left( \frac{\tau^* - t_0}{T} + 1\right) $, the initial condition
\begin{equation}
\hspace*{-0.3cm}
V(x_0) = \left(\frac{\left(\frac{\tau^* -t_0}{T} + 1 \right) }{\sigma_{\kappa} (\tau^* - t_0)\left( c^2_{\kappa,2}\norm{E} - \frac{k c^2_{\kappa,1}}{2} (\frac{\tau^*- t_0}{T} + 1) \right)} \right)^{\frac{1}{\sigma_{\kappa}}}
\label{critical-V_0}
\end{equation}
where $\sigma_{\kappa} = 2q_{\kappa} - 1$, results in the denominator of \eqref{eq:remark_inequality} being zero at time $\tau=\tau^*$. Hence, the function $V(\bar{x}(\tau))$ in \eqref{eq:remark_inequality} has no guaranteed finite upper bound.  
In fact, for any $\norm{E} > k c^2_{\kappa,1} / 2 c^2_{\kappa,2}$, there exists $\tau^*$ sufficiently close to $t_0$, and an associated $V(x_0)$ as in \eqref{critical-V_0}, such that \eqref{eq:remark_inequality} is violated. In conclusion, we can only expect the result of Theorem \ref{it:theorem_part_2} to be provable when simultaneously $\kappa \ge 2$ and $ f(t,x) = E \nabla F(x) $, if the unknown condition $k > (2 c^2_{\kappa,2}) /c^2_{\kappa,1} \norm{E}$ is satisfied.

\textit{6.3. Nonvanishing drift term:}	By Proposition \ref{proposition}, convergence on the average is achieved with the PT seeker even with a nonvanishing bounded perturbation, whereas for the 
asymptotic seeker \eqref{eq:exponential_seeker}, only {\em local} ISS can be established with respect to a bound on the drift.
This highlights another advantage of  the PT Seeker. 

\textit{6.4.~Feasibility:} Finally, the practical implementation of \eqref{eq:closed_loop_system_t} might appear as an issue. From \eqref{eq:u1} and \eqref{eq:u2}, it is clear that the forward and the angular velocities $u_1(t)$ and $u_2(t)$ grow to infinity as the unicycle approaches the source $x^*$. As a workaround, the fractional expressions in \eqref{eq:pt_controlla} can simply be "clipped" to their physically maximal possible values. As a consequence of this, the vehicle is steered to a small neighborhood of the source using a PT Seeker and then being further driven closer to the source using an asymptotic Seeker. In practice, rather than letting $\nu(t-t_0)$ go to zero in \eqref{eq:pt_controlla}, it is sufficient to let it decay down to 0.2--0.3.

	\section{Simulation Results}
	%
		\begin{figure}[t]
			\hspace*{-0.2cm}
			\includegraphics[trim = {2.5cm 0cm 0cm 1cm},clip,scale = 0.29]{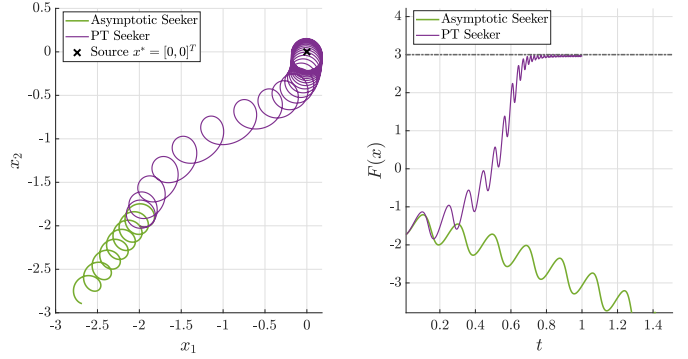}
			\caption{In the presence of the unstable vanishing drift \eqref{eq:drift_term},  the trajectory of the PT Seeker \eqref{eq:closed_loop_system_t} reaches the source at the origin, i.e., the peak $F(x^*) = 3$ of the field \eqref{eq:exp_field} in time $T = 1$, whereas the trajectory of the asymptotic Seeker \eqref{eq:exponential_seeker} moves away from the source $x^*$, i.e., the peak $F(x^*) = 3$.}
			\label{fig:trajectories}
		\end{figure}
		\label{sec:simulation}
		In this section, the application of the PT-control law \eqref{eq:pt_controlla} is demonstrated on examples where Assumption \ref{ass:drift_term} holds. For comparison, the PT Seeker \eqref{eq:closed_loop_system_t} is simulated in pair with its corresponding asymptotic Seeker \eqref{eq:exponential_seeker} using the same parameters. As discussed in Section 6.4, throughout  the simulations we implement the 'clipping' of \eqref{eq:nu} as
		$\tilde{\nu}(t-t_0) = \max\left(0.3, \nu(t-t_0) \right)$. 
		With this, the time-varying gain in  \eqref{eq:pt_controlla} reads as
		\begin{equation}
		\frac{1}{\tilde{\nu}^2(t-t_0)} =  \min\left(\frac{1}{0.3^2}, \frac{1}{\nu^2(t-t_0)}\right) \label{eq:kt}
		\end{equation}
		which is finite as $t \rightarrow t_0 + T$.

	\textit{7.1.~Vanishing drift:}	For the numerical example when Assumption \ref{ass:drift_A} holds, i.e., $f(t,x) = E \nabla F(x)$, the field is chosen as
		\begin{equation}
		F(x) = 2.5 + \frac{1}{2} \cos\left( \frac{2 \pi}{3} x_1 \right) - \frac{1}{2}x^\top x \label{eq:exp_field}
		\end{equation} with a peak of $F(x^*) = 3$ and a source at the origin, i.e., $x^*=[0,0]^\top$.
		With $E = - I$ where  $I \in \R^{2 \times 2}$ is the identity matrix, the vanishing drift term  
		\begin{equation}
		 \left. \frac{\partial E \nabla F(x) }{\partial x} \right|_{x=x^*} = \begin{bmatrix}
		2(\pi^2 / 9 + 1) &0 \\ 0 &1
		\end{bmatrix} \label{eq:drift_term}
		\end{equation}
		is unstable at the origin, due to its positive eigenvalues $\lambda_{1} \approx 4.19 $ and $\lambda_2 =1 $.
		The parameters are set to $t_0 = 0$, $\omega = 30$, $k = 1.8$, $\mu = 0.001$ and the initial condition $x_0 =[-2,-2]^\top$. The prescribed time is chosen to be $T = 1$.
		From Figure \ref{fig:trajectories}, we observe that the PT Seeker \eqref{eq:closed_loop_system_t} reaches a small neighborhood of the source $x^*$ despite the unstable drift term \eqref{eq:drift_term}. Furthermore, the peak of $F(x)$, i.e., $F(x^*) = 3$ is reached in the prescribed time $t = T = 1$. In contrast to that, the asymptotic Seeker \eqref{eq:exponential_seeker} drifts away from the origin $x^*$ and is not able to overcome the influence of the unstable drift term \eqref{eq:drift_term}. This implies that $k$ does not satisfy the convergence condition \eqref{eq:convergence_cond}. 
		\begin{figure}[t]
			\hspace*{-0.2cm}
			\includegraphics[trim = {2.5cm 0cm 0cm 1cm},clip,scale = 0.28]{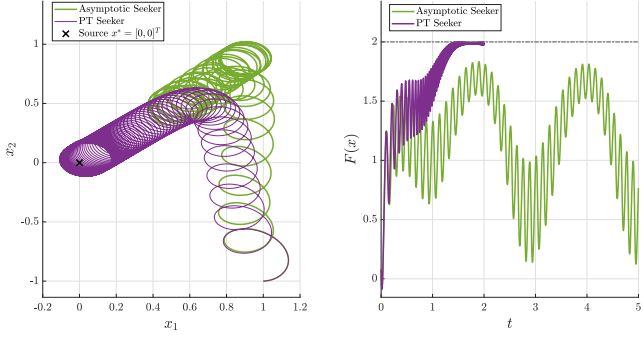}
			\caption{In the presence of the bounded nonvanishing drift \eqref{eq:nonvanish_drift}, the trajectory of the PT Seeker \eqref{eq:closed_loop_system_t} reaches the source $x^*$ at the origin, i.e., the peak $F(x^*) = 2$ of the field \eqref{eq:nonvanish_drift_field_simulation} in time $T = 2$, whereas the trajectory of the asymptotic Seeker \eqref{eq:exponential_seeker} oscillates above the source $x^*$, i.e., below the peak $F(x^*) = 2$.}
			\label{fig:trajectories_bounded}
		\end{figure}
		
	\textit{7.2. Nonvanishing drift:}
		In this case, we choose the signal field as
		 \begin{equation}
		 	F(x) = \cos(x_1) + \cos(x_2) - \frac{1}{2}(x_1^4 + x_2^4). \label{eq:nonvanish_drift_field_simulation}
		 \end{equation}
		 with a source at the origin $x^* = [0,0]^\top$ and a peak of $F(x^*) = 2$.
		The nonvanishing drift term is 
		\begin{equation}
			f(t,x) = (1+\sin(3t)) \cos(x), \label{eq:nonvanish_drift}
		\end{equation} where it is clear that \eqref{eq:nonvanish_drift} satisfies Assumption \ref{ass:drift_B}, since $\norm{f(t,x)} \le \sqrt{2}(1 + \sin(3t)) \le 2 \sqrt{2}$.
		We select the parameters as $t_0 = 0$, $\mu = 0.001$, $\omega = 50$, $k = 1$, $x_0 = [1,-1]^\top$  and $T = 2$.
		As seen in Figure \ref{fig:trajectories_bounded}, the PT Seeker \eqref{eq:closed_loop_system_t} reaches a small neighborhood of the source $x^*$ in spite of the influence of the bounded nonvanishing drift term \eqref{eq:nonvanish_drift}. Moreover, the peak $F(x^*) = 2$ is reached in the prescribed time $T = 2$. Contrary to this, we observe that while the trajectory of the asymptotic Seeker \eqref{eq:exponential_seeker} is bounded, it oscillates above the source $x^*$, i.e., below the peak $F(x^*) = 2$.  This suggests that the nonvanishing (periodic) drift \eqref{eq:nonvanish_drift} cannot be dominated by the asymptotic seeker.

		\begin{figure}[t]
			\hspace*{-0.3cm}
			\includegraphics[trim = {2.5cm 0cm 0cm 0.8cm},clip,scale = 0.28]{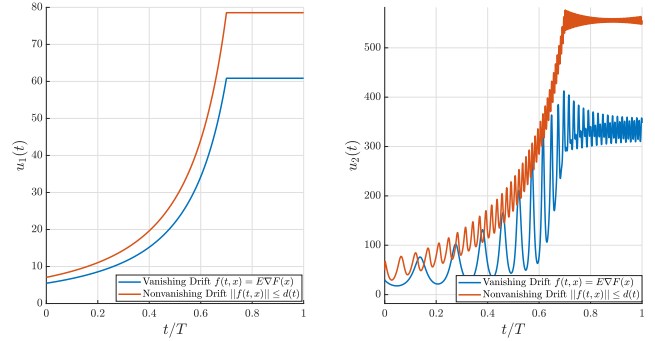}
			\caption{The bounded forward velocity $u_1$ and the angular velocity $u_2$  of the PT Seeker \eqref{eq:closed_loop_system_t} in the case of the vanishing drift \eqref{eq:drift_term} and nonvanishing drift \eqref{eq:nonvanish_drift}, respectively.}
			\label{fig:inputs}
		\end{figure}
	It is worth noting that in both cases the inputs of the PT Seeker remain bounded as due to the clipping \eqref{eq:kt} (see Figure~\ref{fig:inputs}). 
		
\begin{figure}
		\hspace*{-0.5cm}
		\includegraphics[trim = {0cm 0cm  -1cm 1cm},clip,scale = 0.35]{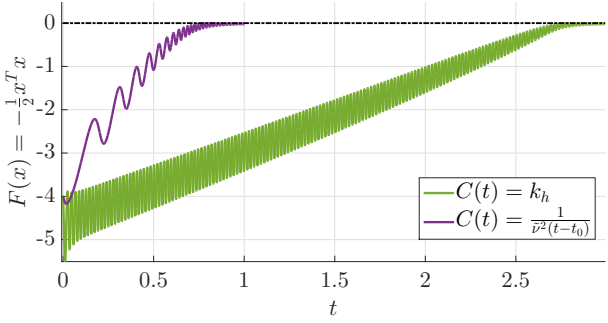}
		\caption{Time elapsed for the saturated seeker \eqref{eq:sat_seeker} to reach the peak of the field $F(x^*) =0$ with the constant gain $k_h$ and the time-varying gain \eqref{eq:kt}.}
		\label{fig:field_sat}
	\end{figure}

\begin{figure}
		\hspace*{-0.3cm}
		\includegraphics[trim = {2.5cm 0cm 0cm 1cm},clip,scale = 0.28]{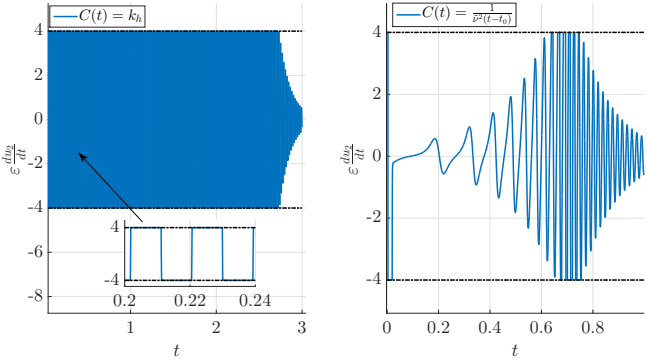}
		\caption{The saturated angular acceleration \eqref{eq:sat_acc} of the seeker \eqref{eq:sat_seeker} with the constant gain $k_h = 11.11$ and the time-varying gain \eqref{eq:kt}.}
		\label{fig:acc_sat}
	\end{figure}

		\textit{7.3. Time-varying gain vs constant high gain:}
		\label{sec:gains_debate}
		From the above simulations, it is clear that the bounded implementation \eqref{eq:kt} of the time-varying gain in \eqref{eq:pt_controlla} is sufficient for the PT Seeker to achieve practical PT convergence to the source $x^*$, despite the influence of the drift $f(t,x)$.
		However, one could argue that the same can be achieved if the control law \eqref{eq:pt_controlla} is multiplied with a sufficiently high constant gain. Hence, it is natural to ask the following question. What is the advantage of using a  bounded time-varying gain over a constant high gain in the control law \eqref{eq:pt_controlla}?
		To answer this, let us compare both cases through a simulation. Consider a drift free scenario where the angular velocity of the unicycle can only be accessed through its angular acceleration which in addition is saturated. The angular velocity tuning control law \eqref{eq:pt_controlla} can be realized through actuating the saturated angular acceleration as
		\begin{subequations}
			\label{eq:sat_seeker}
			\begin{align}
			\dot{x}  &= 
			\underbrace{C(t) \sqrt{\omega}}_{u_1} \begin{bmatrix}
			\cos(\theta) \\
			\sin(\theta)
			\end{bmatrix}
			\\
			\dot{\theta}   &= u_2  \\ 
			\dot{u}_2 &= \frac{1}{\epsilon}  \text{sat}\left(-u_2 + C(t)\left( \omega - \frac{k}{\mu}(y-z) \right)\right) \label{eq:sat_acc} \\
			\dot{z}   &=  C(t)\frac{1}{\mu}(y-z). 
			\end{align}
		\end{subequations}
		where $\epsilon > 0$ and the saturation function with the saturation level $S >0$ is defined as $\text{sat}(x) = \min(S,\max(-S,x))$.
		Observing the steady state $u_2^*$ of \eqref{eq:sat_acc}, it becomes clear that it is equivalent to  \eqref{eq:u2} which implies that the angular velocity tuning can be achieved through the angular acceleration \eqref{eq:sat_acc} for $\epsilon \ll \mu$.
		The factor $C(t)$ is used to differentiate between the time-varying gain and the constant high gain.
		Namely, we simulate the seeker \eqref{eq:sat_seeker} for the case where $C(t)$ is set to \eqref{eq:kt} and $C(t)$ is set to some constant gain $k_h > 0$ and compare the results. Since the maximal value of \eqref{eq:kt} for $t_0 = 0$ and $T  = 1$ is $1/0.3^2 = 11.11$, the constant gain is chosen as $k_h = 11.11$. 
		For the simulations, the parameters are selected as $t_0 = 0$, $\epsilon = 0.0004$, $\mu = 0.001$, $\omega = 30$, $k = 1.4$ and the initial condition as $x_0 = [2,2]^\top$. The prescribed time is $T = 1$ and the saturation level is $S = 4$. The signal field is chosen as $F(x) = -\frac{1}{2} x^\top x$ with a 
		source at the origin, i.e.,  $x^* = [0, 0]^\top$ and a peak $F(x^*) = 0$. 		
		Inspecting Figure \ref{fig:field_sat}, we observe that the seeker \eqref{eq:sat_seeker} reaches the source $x^*$ at the origin, i.e., the peak $F(x^*) = 0$ with both the time-varying gain \eqref{eq:kt} and the constant gain $k_h = 11.11$. However, while the seeker \eqref{eq:sat_seeker} with the time-varying gain \eqref{eq:kt} reaches the peak $F(x^*) = 0$ in the prescribed time $t = T$, the seeker with the constant gain $k_h = 11.11$  reaches the peak in a time that is approximately three times longer than $T$. 
		The reason for this is best explained by examining Figure \ref{fig:acc_sat}. Herein, we observe the angular acceleration \eqref{eq:sat_acc} saturated at $\pm 4$. Additionally, the control effort for \eqref{eq:sat_acc} with the constant gain $k_h = 11.11$ is greater than the control effort with the time-varying gain \eqref{eq:kt}. Despite this, the convergence speed of the seeker \eqref{eq:sat_seeker} with $C(t) = k_h$ is slower than the case when $C(t)$ is \eqref{eq:kt} due to the saturation limits on the angular acceleration.  
		This implies that a time-varying gain retains the PT convergence to the source $x^*$ while simultaneously keeping the control effort moderately low. Therefore, we can conclude that a time-varying gain is advantageous over a constant high gain in scenarios where there is limited actuation.

	\section{Conclusion}
	 We develop a  source seeking algorithm (PT Seeker) that is able to converge to a small, but bounded neighborhood of the source in user prescribed time regardless of the initial conditions. For this, we use the temporal transformation approach and adapt the seeking scheme in \cite{ES_using_LBAandSP}, achieving the newly introduced FxT-sSPUAS convergence. This results in scaling the velocities of the unicycle with time-varying gains that grow unbounded as the terminal time is approached. In addition to PT convergence, these time-varying gains provide the PT Seeker \eqref{eq:closed_loop_system_t} with robustness to bounded external forces and possibly destabilizing gradient dependent drifts. For the implementation, the growth of these unbounded gains is stopped at user-defined maximum values. Thus, the control inputs remain bounded which keeps the control effort moderately low, but at the same time the PT convergence  and the robustness to the drift are \textit{practically} retained. We confirm our theoretical results by simulations and show the advantages of  our method over merely asymptotic source seeking. 
	



	\section*{Appendix}
	Consider the coupled system \eqref{eq:general_nominal_system} and assume existence of a quasi-steady state $l: \R^n \rightarrow \R^m $. Let $\mathcal{S} \subseteq \R^n$ be a compact set. The solution of a differential equation $\frac{\text{d} x}{\text{d} t} = f(x, t) $ with $x(t_0) = x_0$ is represented by $x( \cdot, t_0, x_0): \R \rightarrow \R^n $.

	
	\begin{definition}
		The set $\mathcal{S}$ is said to be \textbf{fixed-time singularly practically uniformly stable} for \eqref{eq:general_nominal_system} if for all $T>0$ and 
for all $\epsilon_x, \epsilon_z \in (0, \infty)$ there exist $\delta_x, \delta_z \in (0,\infty)$ and $\omega_0 \in (0,\infty)$ such that for all $\omega \in (\omega_0,\infty)$ there exists a $\mu_0 \in (0, \infty)$ such that for all $\mu \in (0,  \mu_0 )$ and for all $t_0 \geq 0$ 
		\begin{equation*}
		\begin{aligned}
		&x_0 \in U_{\delta_x}^{\mathcal{S}}  \quad {\rm and} \quad z_0 - l(x_0) \in  U_{\delta_z}^{0} \implies  x\left(t,t_0,x_0 \right) \in U_{\epsilon_x}^{\mathcal{S}}  \\
		&{\rm and} \quad z\left(t, t_0, x_0 \right) - l\left(x\left(t,t_0,x_0\right)\right) \in U_{\epsilon_z}^{0}, ~t \in [t_0, t_0 + T). 
		\end{aligned}
		\end{equation*}
	\end{definition}
	
	\begin{definition}\label{def:FxT-sPUA}
		The set $\mathcal{S}$ is said to be \textbf{fixed-time singularly practically uniformly attractive (FxT-sPUA)} for \eqref{eq:general_nominal_system} if for all $T>0$ and 
for all $\delta_x, \delta_z, \epsilon_x, \epsilon_z \in (0, \infty)$ there exist a $\tau_f \in [0, T)$ and $\omega_0 \in (0,\infty)$ such that for all $\omega \in (\omega_0,\infty)$ there exists a $\mu_0 \in (0, \infty)$ such that for all $\mu \in (0,  \mu_0 )$ and for all $t_0 \geq 0$
	\begin{equation*}
		\begin{aligned}
		&x_0 \in U_{\delta_x}^{\mathcal{S}}  \quad {\rm and} \quad z_0 - l(x_0) \in  U_{\delta_z}^{0} \implies \nonumber \\
		&x\left(t, t_0, x_0 \right)\in U_{\epsilon_x}^{\mathcal{S}},  ~t \in \left[ t_0 + \frac{\tau_f/\mu}{1 + \frac{\tau_f}{\mu T}}, t_0 + T \right)  {\rm and}  \nonumber \\
		&z\left(t, t_0, x_0 \right) - l\left(x\left(t,t_0,x_0\right)\right) \in U_{\epsilon_z}^{0},  ~t \in \left[ t_0 + \frac{\tau_f}{1 + \frac{\tau_f}{T}}, t_0 + T \right)    \nonumber
		\end{aligned}
		\end{equation*}
	\end{definition}
	
	\begin{definition}
		The solutions of \eqref{eq:general_nominal_system} are said to be \textbf{fixed-time singularly practically uniformly bounded with respect to $\mathcal{S}$} if for all $T>0$ and 
  for all $\delta_x, \delta_z \in (0,\infty)$  there exist $\epsilon_x, \epsilon_z \in (0, \infty)$  and $\omega_0 \in (0,\infty)$ such that for all $\omega \in (\omega_0,\infty)$ there exists a $\mu_0 \in (0, \infty)$ such that for all $\mu \in (0,  \mu_0 )$ and for all $t_0 \geq 0$ 
		\begin{equation*}
		\begin{aligned}
		&x_0 \in U_{\delta_x}^{\mathcal{S}}  \quad {\rm and} \quad z_0 - l(x_0) \in  U_{\delta_z}^{0} \implies  
		x\left(t,t_0,x_0 \right) \in U_{\epsilon_x}^{\mathcal{S}} \\ &{\rm and} \quad z\left(t, t_0, x_0 \right) - l\left(x\left(t,t_0,x_0\right)\right) \in U_{\epsilon_z}^{0}, ~t \in [t_0, t_0 + T). 
		\end{aligned}
		\end{equation*}
	\end{definition}
	
	\begin{definition}\label{def:FxT-sSPUAS}
		The set $\mathcal{S}$ is said to be \textbf{fixed-time singularly semi-globally practically uniformly asymptotically stable (FxT-sSPUAS)} for \eqref{eq:general_nominal_system} if the set $\mathcal{S}$ is fixed-time singularly practically uniformly stable, fixed-time singularly practically uniformly attractive and the solutions of \eqref{eq:general_nominal_system} are fixed-time singularly practically uniformly bounded with respect to $\mathcal{S}$.
	\end{definition}
The above FxT-sSPUAS definitions correspond to the sSPUAS definitions in \cite{ES_using_LBAandSP}, but adjusted for the finite time interval $I_t$ using  \eqref{eq:tau_to_t} and \eqref{eq:t_to_tau} and their equivalence is seen by letting $T \rightarrow \infty$.
%
	
	
\bibliography{references}

\subsection*{}

\setlength\intextsep{0pt} 
\begin{wrapfigure}{l}{0.13\textwidth}
	\centering
	\includegraphics[width=0.15\textwidth]{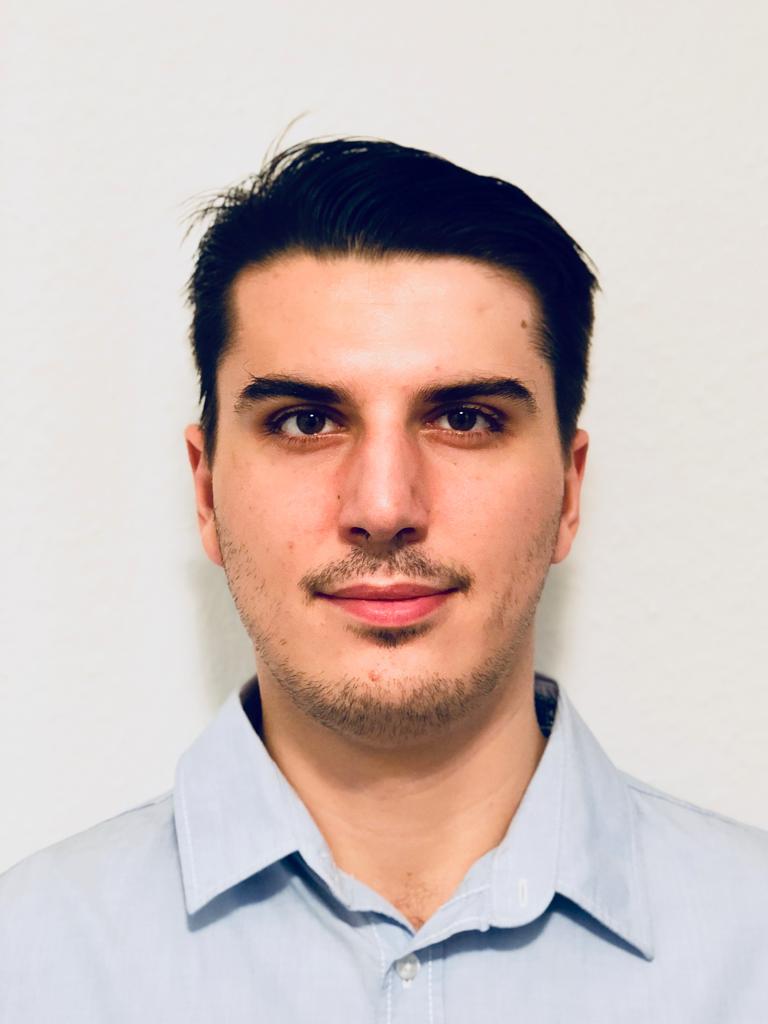}
\end{wrapfigure}
\noindent \textbf{\bf Velimir Todorovski} received M.Sc. degree in Electrical and Computer Engineering with focus on Automation and Robotics from Technical University of Munich, Munich, Germany in 2022. He is currently pursuing a Ph.D.
degree at the Chair of Information-Oriented Control at Technical University of Munich. Some of
his research interests include data-driven control with safety and performance guarantees, prescribed-time control, source seeking, model-free stabilization and fault diagnosis.

\subsection*{}

\setlength\intextsep{0pt} 
\begin{wrapfigure}{l}{0.13\textwidth}
	\centering
	\includegraphics[width=0.15\textwidth]{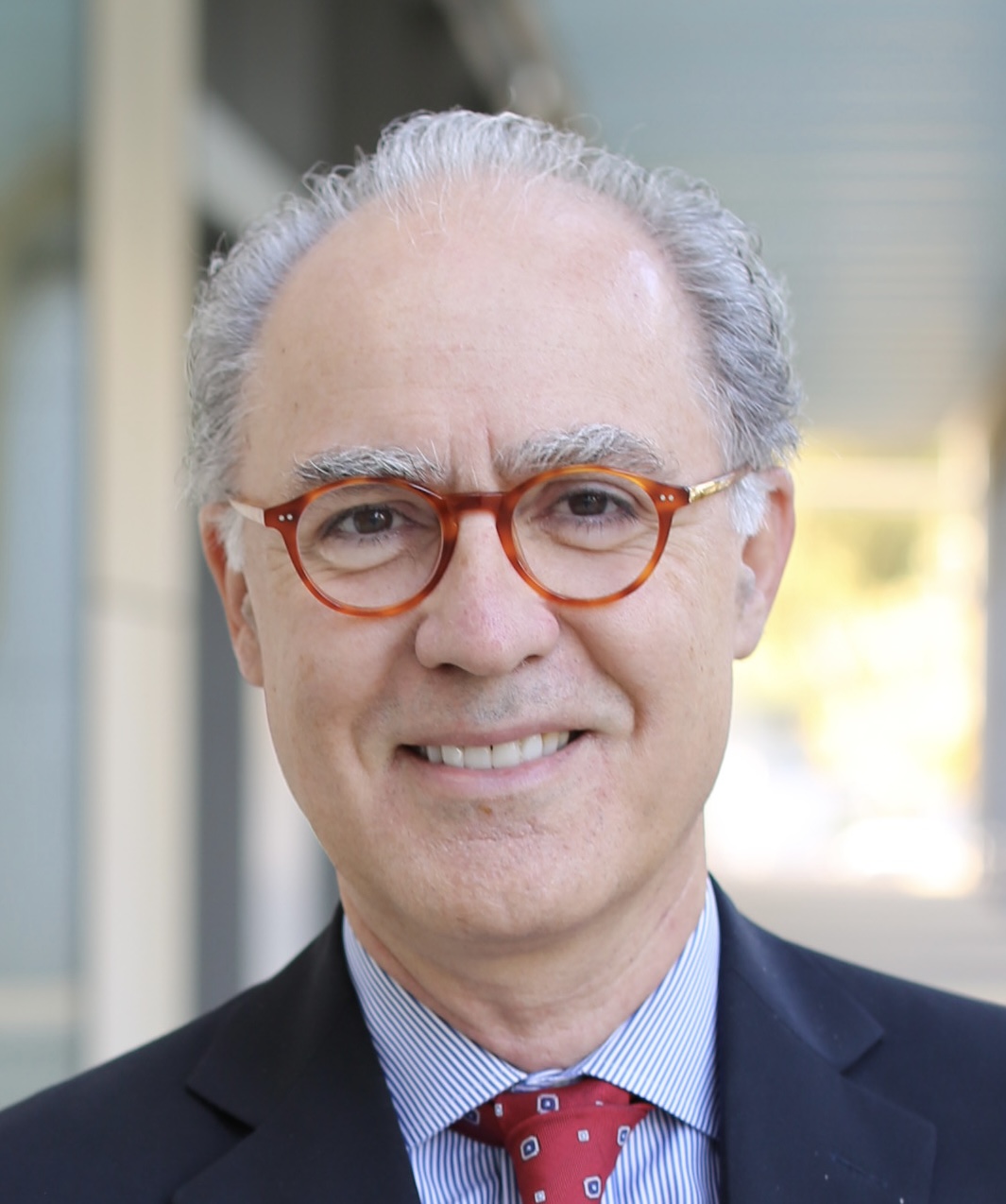}
\end{wrapfigure}
\noindent \textbf{\bf Miroslav Krstic} is Distinguished Professor of Mechanical and Aerospace Engineering, holds the Alspach endowed chair, and is the founding director of the Cymer Center for Control Systems and Dynamics at UC San Diego. He also serves as Senior Associate Vice Chancellor for Research at UCSD. As a graduate student, Krstic won the UC Santa Barbara best dissertation award and student best paper awards at CDC and ACC. Krstic has been elected Fellow of seven scientific societies - IEEE, IFAC, ASME, SIAM, AAAS, IET (UK), and AIAA (Assoc. Fellow) - and as a foreign member of the Serbian Academy of Sciences and Arts and of the Academy of Engineering of Serbia. He has received the Richard E. Bellman Control Heritage Award, SIAM Reid Prize, ASME Oldenburger Medal, Nyquist Lecture Prize, Paynter Outstanding Investigator Award, Ragazzini Education Award, IFAC Ruth Curtain Distributed Parameter Systems Award, IFAC Nonlinear Control Systems Award, Chestnut textbook prize, Control Systems Society Distinguished Member Award, the PECASE, NSF Career, and ONR Young Investigator awards, the Schuck (’96 and ’19) and Axelby paper prizes, and the first UCSD Research Award given to an engineer. Krstic has also been awarded the Springer Visiting Professorship at UC Berkeley, the Distinguished Visiting Fellowship of the Royal Academy of Engineering, the Invitation Fellowship of the Japan Society for the Promotion of Science, and four honorary professorships outside of the United States. He serves as Editor-in-Chief of Systems \& Control Letters and has been serving as Senior Editor in Automatica and IEEE Transactions on Automatic Control, as editor of two Springer book series, and has served as Vice President for Technical Activities of the IEEE Control Systems Society and as chair of the IEEE CSS Fellow Committee. Krstic has coauthored eighteen books on adaptive, nonlinear, and stochastic control, extremum seeking, control of PDE systems including turbulent flows, and control of delay systems.

\end{document}